\documentclass[12pt]{iopart}
\usepackage{graphicx}
\usepackage[english]{babel}
\usepackage{cite}
\usepackage{url}

\newcommand{\n}[1]{\ensuremath{|\mathbf{#1}|}}
\newcommand{\nuC}{\ensuremath{^{12}_{\phantom{1}6}\textrm{C}(\nu_\mu,\mu^-)}}
\newcommand{\nuCX}{\ensuremath{^{12}_{\phantom{1}6}\textrm{C}(\nu_\mu,\mu^-)X}}
\newcommand{\ph}{\ensuremath{2p2h}}
\newcommand{\genie}{\textsc{genie}}
\newcommand{\vt}{\textsc{genie}$+\nu$T}
\newcommand{\globes}{\textsc{gl}{\small o}\textsc{bes}}

\begin{document}

\title{Systematic uncertainties in long-baseline neutrino-oscillation experiments}

\author{Artur M Ankowski and Camillo Mariani\footnote{Emerging Leader}}
\address{Center for Neutrino Physics, Virginia Tech, Blacksburg, Virginia 24061, USA}
\vspace{10pt}

\begin{abstract}
Future neutrino-oscillation experiments are expected to bring definite answers to the questions of neutrino-mass hierarchy and violation of charge-parity symmetry in the lepton sector. To realize this ambitious program it is necessary to ensure a significant reduction of uncertainties, particularly those related to neutrino-energy reconstruction.
In this review, we discuss different sources of systematic uncertainties, paying special attention to those arising from nuclear effects and detector response. By analyzing nuclear effects we show the importance of developing accurate theoretical models, capable to provide quantitative description of neutrino cross sections, together with the relevance of their implementation in Monte Carlo generators and extensive testing against lepton-scattering data. We also point out the fundamental role of efforts aiming to determine detector responses in test-beam exposures.
\end{abstract}

%
%
\submitto{\jpg}
%
%
%

\section{Introduction}
Neutrinos are produced in states of given flavors---electron, muon or tau---that are mixtures of mass eigenstates. As different masses propagate with different phases, the mixture at some distance from a neutrino source may be different from the initial one, resulting in a disappearance of the initial flavor and an appearance of other flavors in the neutrino beam.

Quantitatively, the phase of the state of mass $m_i$ and momentum ${\bf p}$ at the time $t$ from the production is \[
t\sqrt{{\bf p}^2+m_i^2}\approx\n p t+{m_i^2L/2E_\nu},\]
$E_\nu$ being the neutrino energy and $L$ denoting the distance from the source. Therefore, the disappearance and appearance probabilities depend on the squared mass differences, $\Delta m_{ij}^2=m_{i}^2-m_{j}^2$, ratio $L/E_\nu$, and mixing angles, $\theta_{ij}$, describing how mass eigenstates mix to form states of definite flavor.

This phenomenon, called the neutrino oscillations, has been well established in the past two decades and is now considered one of the major discoveries in particle physics, celebrated by the recent Nobel Prize~\cite{Kajita:2016cak,McDonald:2016ixn}. The values of the oscillation parameters---two squared mass differences and three mixing angles---are being measured with an increasing precision, thanks to a global effort that produced a number of experiments currently taking data or planned in the near future.

Within the next two decades,  the precision is expected to be sufficient to begin testing the unitarity of the mixing matrix~\cite{Kim:2014rfa,An:2015jdp}, discover the neutrino-mass hierarchy~\cite{Patterson:2015xja} and unambiguously determine the value of the Dirac phase, $\delta_{CP}$, violating the charge-parity (CP) symmetry of neutrino mixing~\cite{Abe:2015zbg,Acciarri:2015uup,1499045}. These results are going to have profound consequences for possible extensions of the Standard Model, building models of the neutrino masses and our understanding of the matter-antimatter asymmetry in the Universe.

In this review, we discuss systematic uncertainties in ongoing and future long-baseline ($\sim$300--1300 km) neutrino-oscillation experiments using conventional beams~\cite{Abe:2015zbg,1499045,Abe:2011ks,Ayres:2007tu,Acciarri:2016crz}. Such neutrino beams are tertiary products, originating from the decay of mesons---predominantly pions---produced in interactions
of the primary proton beam with a target. As the resulting neutrinos are not monoenergetic, to extract the oscillation parameters from the collected neutrino-event distribution, their energies have to be reconstructed on an event-by-event basis from the measured kinematics of particles in the final state. Depending on the baseline, the relevant neutrino energies extend from a few hundred MeV to a few GeV, at which the dominant interaction mechanisms change from quasielastic (QE) to resonant and nonresonant meson production~\cite{Formaggio:2013kya}.

Because the oscillation parameters are extracted from the energy dependence of event distributions, their accurate determination requires an accurate reconstruction of neutrino energy. While the reconstruction would involve smallest uncertainties for scattering off free protons and deuteron~\cite{Moreno:2015nsa}, owing to the low cross sections involved, it is necessary to employ nuclear targets as detector materials to ensure high statistics of collected events. As a consequence, however, description of nuclear effects turns out to be one of the largest sources of systematic
uncertainties in the oscillation analysis of modern long-baseline experiments. Additionally, when energy reconstruction requires the kinematics of produced hadrons to be measured, detector effects may play an important role and uncertainties of the detector response contribute to the systematic uncertainties of the extracted oscillation parameters~\cite{Ankowski:2015jya,Ankowski:2015kya}. In particular, neutrons typically travel some distance from the primary interaction vertex before depositing part of their energy in the detector, which makes them problematic to associate with the neutrino event and leads to an underestimation of the neutrino energy.

Over the past decade, the extensive neutrino-scattering program has yielded a wealth of experimental cross sections for carbon or hydrocarbon, CH~\cite{Wu:2007ab,Lyubushkin:2008pe,AguilarArevalo:2010zc,AguilarArevalo:2010cx,Nakajima:2010fp, AguilarArevalo:2010bm,AguilarArevalo:2010xt,Fields:2013zhk,Fiorentini:2013ezn,Abe:2013jth,Aguilar-Arevalo:2013dva, Abe:2014nox,Tice:2014pgu,Higuera:2014azj,Aguilar-Arevalo:2013nkf,Walton:2014esl,Abe:2015oar,Eberly:2014mra,Aliaga:2015wva, Abe:2014iza,Rodrigues:2015hik,Mousseau:2016snl,Abe:2016tmq,Marshall:2016rrn,McGivern:2016bwh,Abe:2016fic,DeVan:2016rkm,Marshall:2016yho}, and much fewer results for other targets, such as water\cite{Abe:2014dyd,Abe:2016aoo}, argon \cite{Anderson:2011ce,Acciarri:2014isz,Acciarri:2014eit}, iron \cite{Tice:2014pgu,Abe:2014nox,Mousseau:2016snl,Tzanov:2005kr,Adamson:2009ju,Abe:2015biq,Adamson:2016hyz} and lead \cite{Tice:2014pgu,Mousseau:2016snl}. It is important to note that there seem to remain puzzling tensions between different measurements---such as \cite{Lyubushkin:2008pe,AguilarArevalo:2010zc,Fiorentini:2013ezn} or \cite{AguilarArevalo:2010bm,Eberly:2014mra}---which have attracted a sizable attention of the theoretical community. A number of approaches developed to
describe nuclear response to other probes---such as electrons and pions---has been extended to neutrino interactions, and the understanding of nuclear effects relevant to neutrino-oscillation experiments has clearly improved~\cite{Benhar:2009wi,Leitner:2010kp,Martini:2012fa,Nieves:2012yz,Meloni:2012fq,Lalakulich:2012hs,Martini:2012uc,Coloma:2013rqa, Coloma:2013tba,Mosel:2013fxa,Jen:2014aja,Ankowski:2014yfa,Ericson:2016yjn,Ankowski:2016bji}; see also recent reviews~\cite{Alvarez-Ruso:2014bla,Benhar:2015wva}. However, many problems still await a quantitative explanation. While for detector targets containing hydrogen a method to separate out the pion-production events on free protons has been proposed~\cite{Lu:2015hea}, current and future neutrino-oscillation experiments are going to collect data predominantly for interactions with the targets of atomic numbers $A$ ranging from 12 to 40, for which accurate and complete nuclear models
permitting fully trustable data analysis are not available yet.

The main difficulty of developing theoretical models useful to oscillation experi{\-}ments stems from the flux average over polychromatic beams~\cite{Benhar:2010nx}. As any data point may receive contributions from a range of neutrino energies and different interaction mechanisms, for a theoretical model to be useful at the kinematics of long-baseline oscillation experiments, it has to cover broad kinematic region, describe multiple scattering processes~\cite{Benhar:2015wva} and be able to describe relativistic products of interaction. Furthermore, experiments employing tracking detectors require predictions of exclusive cross sections for different hadronic final states.

In this review we pay particular attention to neutrino energies in the 1-GeV region because of its relevance for the next generation of oscillation experiments. Note that the narrow beam of the Hyper-Kamiokande experiment is going to be peaked at 0.6~GeV~\cite{Abe:2015zbg}, and in Deep Underground Neutrino Experiment (DUNE)~\cite{Acciarri:2015uup} the region of the second oscillation maximum---much more sensitive to the Dirac phase than the first one---is going to correspond to $\sim$0.8 GeV. Having in mind that due to much larger statistics the disappearance channels are going to be more sensitive to systematic uncertainties, we usually consider as a case study a disappearance experiment of the setup similar to that of T2K~\cite{Abe:2011ks}, but highly idealized. At this kinematics, QE scattering is the main interaction mechanism and its uncertainties play the most important role. Note that currently the cross section for QE scattering induced by two-body currents is determined from the near-detector data after subtracting the dominant QE contribution induced by one-body current and other contributions. As a consequence, the uncertainties related to one-body QE scattering are of utmost importance. Discussing them in details, we point out the relevance of realistic description of the ground-state properties of the target nucleus. We argue that
employing the same target in the near and far detectors can be an effective method for reducing systematic uncertainties in any interaction channel. We also emphasize the importance of new cross section measurements, developing more accurate nuclear models and improving existing Monte Carlo generators. In addition, we discuss the role of detector effects, which for the calorimetric method of neutrino-energy reconstruction are of comparable importance as the uncertainties of two-body QE scattering. Analyzing the influence of missing energy on $\delta_{CP}$ measurement in an experiment similar to DUNE, we illustrate the significance of determining the detector response in test-beams exposures.

After general remarks regarding sys{\-}te{\-}ma{\-}tic uncertainties of measurements of CP violation in \sref{sec:generalRemarks}, we present our method of the oscillation analysis of mock data in \sref{sec:oscAnalysis}.
In sections \ref{sec:nuclearEffects} and \ref{sec:detectorEffects} we illustrate the importance of nuclear and detector effects for an accurate reconstruction of neutrino energy and an unbiased extraction of the oscillation parameters, discussing selected results from publications~\cite{Ankowski:2015jya,Ankowski:2015kya,Coloma:2013tba,Jen:2014aja,Ankowski:2016bji}. Finally in \sref{sec:summary} we summarize this review.

\section{General remarks}\label{sec:generalRemarks}
In recent and ongoing oscillation experiments, it has been a common practice to use the spread between different theoretical descriptions of nuclear effects or different implementations of the same model in Monte Carlo generators as an estimate of the associated systematic uncertainties. While this method seems very effective, a word of caution is in order. Non-negligible correlations between the ingredients used to evaluate the uncertainties---such as common assumptions of nuclear models or fine-tuning of event generators to the same data---may lead to sizable underestimates, making this procedure insufficient and unreliable.

For the next 20 years, the long-baseline neutrino-oscillation program is most likely going to employ conventional beams~\cite{Abe:2011ks,1499045,Ayres:2007tu,Abe:2015zbg,Acciarri:2016crz}. This technique of producing neutrino beams has been used for a few decades: an intense proton beam is impinged on a target typically made of beryllium or graphite in order to produce mesons, mainly pions. The target is embedded within a horn---an electromagnet producing a toroidal magnetic field---that focuses secondary particles of a~selected charge and defocuses those of the opposite charge. The horn polarization determines whether the resulting beam will be made of neutrinos or antineutrinos.
It is important to point out that the contamination by wrong-sign mesons introduces much more significant neutrino backgrounds in the antineutrino beams than antineutrino backgrounds in the neutrino beams.

An accurate determination of the neutrino flux---its flavor composition and energy spectrum---exclusively from the proton-beam parameters, meson-production data for thick targets and horn configuration is a challenging task.
State-of-the-art methods of controlling systematic uncertainties of a neutrino beam have been developed for the MINOS~\cite{Michael:2008bc} and MINERvA~\cite{Aliaga:2013uqz} experiments, largely thanks to using two horns of adjustable positions at the NuMI beamline~\cite{Abramov:2001nr}. It is not clear whether such a feature---difficult to realize for off-axis beams---is going to be available for future oscillation experiments. Therefore, the beam-normalization uncertainty of about 5\% achieved in the MINOS and MINERvA experiments, seems to be a reasonable estimate for future oscillation experiments employing conventional beams. Note, however, that in the case of the off-axis T2K beam this uncertainty has been reduced to 3.6\% in the neutrino mode and to 3.8\% in the antineutrino mode, thanks to extensive use of the near-detector data~\cite{Duffy}. Moreover, recently proposed new concepts of neutrino beams, such as an electron and muon neutrino beam from the decay of a stored muon beam (nuSTORM)~\cite{Adey:2014rfv}, would allow the neutrino flux to be determined with an accuracy of the order of 1\%, with the same systematic uncertainties of the neutrino and antineutrino fluxes.


To extract the probability of oscillation between the neutrino flavors $\alpha$ and $\beta$, experiments collect event distributions $R_{\alpha\rightarrow\beta}(\mathcal X)$ with respect to a set of observables $\mathcal X$, such as the charged lepton's energy and cosine of its production angle,
\begin{equation}\label{eq:rate}
R_{\alpha\rightarrow\beta}(\mathcal X)=\mathcal N\int \rmd E_\nu\Phi_\alpha(E_\nu)P(\nu_\alpha\rightarrow\nu_\beta)\frac{\rmd\sigma_\beta}{\rmd\mathcal X}\epsilon_\beta(\mathcal X).
\end{equation}
In the above equation, the normalization factor $\mathcal N$ depends on the beam power, data-collecting time, fiducial mass etc. The flux expected in the detector, $\Phi_\alpha(E_\nu)$, and the oscillation probability $P(\nu_\alpha\rightarrow\nu_\beta)$ are both functions of the true neutrino energy $E_\nu$. The differential cross section ${\rmd\sigma_\beta}/{\rmd\mathcal X}$ describes likelihood for a neutrino of the flavor $\beta$ and energy $E_\nu$ to produce an event of kinematics $\mathcal X$. The detection efficiency is denoted as $\epsilon_\beta(\mathcal X)$.

Systematic uncertainties in neutrino-oscillation experi{\-}ments can be effectively reduced by using an unoscillated event distribution measured in the near-detector system to predict the distribution expected in the far detector. This method of exploiting the cancellation of correlated uncertainties---giving the maximal cancellation when the near and far detectors are functionally identical---has been used with great success in the reactor experiments Daya Bay~\cite{An:2015rpe}, RENO~\cite{RENO:2015ksa} and Double Chooz~\cite{Abe:2014bwa} to measure the mixing angle $\theta_{13}$ with remarkably high precision. Barring the differences in the backgrounds and in the geometric acceptances of the beam, the ratio of the energy-unfolded event distributions in these disappearance experiments, $R_{\alpha\rightarrow\alpha}(E_\nu)$, has the simple interpretation
\begin{equation}\label{eq:event_rate_far_near}
\frac{R_{\alpha\rightarrow\alpha}^\mathrm{far}(E_\nu)}{R_{\alpha\rightarrow\alpha}^\mathrm{near}(E_\nu)}
\approx\frac{{\mathcal N}_\mathrm{far}\Phi_\alpha^\mathrm{far}(E_\nu)
  P(\nu_\alpha\rightarrow\nu_\alpha)}{{\mathcal N}_\mathrm{near}
  \Phi^\mathrm{near}_\alpha(E_\nu) }
=\frac{{\mathcal N}_\mathrm{far}L_\mathrm{far}^2}{{\mathcal N}_\mathrm{near}L_\mathrm{near}^2}\,P(\nu_\alpha\rightarrow\nu_\alpha),
\end{equation}
thanks to small uncertainties of the cross section and well-known relation between $\mathcal X$ and $E_\nu$.
In the above equation, the far-to-near flux ratio, $\Phi_\alpha^\mathrm{far}(E_\nu)/\Phi_\alpha^\mathrm{near}(E_\nu)$, reduces to the ratio of the squared distances from the neutrino source, $L_\mathrm{far}^2/L_\mathrm{near}^2$ and the oscillation probability in the near detector is $P(\nu_\alpha\rightarrow\nu_\alpha)=1$.

Unlike in the reactor experiments, in long-baseline studies, the relevant cross sections are currently known with an accuracy of 10--20\% and the procedure of energy unfolding is much more involved, due to much more complicated relation between the observed event kinematics and neutrino energy. Additionally, the differences in backgrounds and beam acceptances between the near and far sites are non-negligible. These factors altogether largely invalidate the relation~\eref{eq:event_rate_far_near} in this case.



Generally, in $\nu_\mu\rightarrow\nu_e$ appearance measurements, the situation is even more involved. The final and initial neutrino flavors are different, which leads to a dependence of the near-to-far event-distributions ratio on the cross-sections ratio $\sigma_e/\sigma_\mu$. Although (standard-model) theoretical calculations for muon and electron neutrinos are directly related, this may not be the case when the nuclear model employed in the oscillation analysis contains \emph{ad hoc} modifications to reproduce the $\nu_\mu$ event distributions in the near detector. Fortunately, at the kinematics of long-baseline experiments, the charged-lepton mass does not affect the cross
sections very significantly, making the $\sigma_e/\sigma_\mu$ ratio not sizably different from 1.

While the $\sigma_e/\sigma_\mu$ ratio could, in principle, be affected by the second-class currents or deviations from the standard parametrization of the pseudoscalar form factor, it has been recently shown for scattering off free nucleons that only the second-class contribution to the vector current may be relevant at neutrino energies above $\sim$200 MeV~\cite{Day:2012gb}. Subsequently, a similar analysis has been also performed for different implementations of the Fermi gas model~\cite{Akbar:2015yda}. We observe, however, that a nonvanishing second-class contribution would violate conservation of the vector current, invalidating the method to express the vector form factors and to calculate the cross section in~\cite{Day:2012gb,Akbar:2015yda}.

As conventional beams are designed to minimize the $\nu_e$ contamination in order to reduce the background in the far detector, direct measurements of the $\nu_e$ cross sections suffer from the statistics lower typically by 2--3 orders of magnitude than in the $\nu_\mu$ case and from larger uncertainties coming from both the flux and detector response~\cite{Abe:2014usb,Abe:2014agb,Abe:2015mxf,Wolcott:2015hda}. In addition, measurements for electron antineutrinos are even more challenging, due to the lower cross sections involved. Note, however, that a novel method to determine the $\nu_e$ fluxes with uncertainties as low as 1\% has been recently proposed~\cite{Longhin:2014yta}, based on monitoring of kaon decays which are the source of electron neutrinos in conventional neutrino beams.

While determination of $\nu_e$ cross sections could be of vital importance in the presence of nonstandard interactions~\cite{Miranda:2015dra}, so far there is no evidence of their existence \cite{Abe:2014agb,Wolcott:2015hda}. Assuming only standard
interactions and exploiting cancellation of correlated uncertainties between the $\nu_e$ and $\nu_\mu$ channels, the expected number of electron neutrino interactions at far detectors can currently be predicted with $\sim$5\% uncertainty~\cite{Abe:2017uxa}, owing to the uncertainty of the $\sigma_e/\sigma_\mu$ ratio.

While far detectors are designed to maximize the statistics of signal events, the designs of near detectors in long-baseline experiments aim to provide much more detailed information on event kinematics, necessary to reduce uncertainties related to fluxes, interaction models and backgrounds. As a consequence, the near and far detectors frequently differ at the qualitative level~\cite{Michael:2008bc,Abe:2011ks,Ayres:2007tu}. As an illustrative example of the importance of near detectors, we refer to the T2K experiment~\cite{Amaudruz:2012agx}. Using the information from an off-axis magnetized near detector to determine a number of beam-flux and cross-section parameters together with their covariance, the T2K Collaboration is able to reduce the corresponding uncertainties in the far detector from 21.6\% to 2.7\% in the $\nu_\mu\rightarrow\nu_\mu$ disappearance channel~\cite{Abe:2014ugx} and from 25.9\% to 2.9\% in the $\nu_\mu\rightarrow\nu_e$ appearance channel~\cite{Abe:2013hdq}. Nevertheless, the reduction of uncertainties is limited owing to differences between the near and far detectors in target material, angular acceptance and flux~\cite{Quilain:2016ymh}. To allow further progress in reducing the nuclear-model uncertainties, a new near detector employing water scintillator has been recently proposed~\cite{Ovsiannikova:2016add}, expected to determine the ratio of cross-sections between water and hydrocarbon with 3\% uncertainty.

Aiming to determine the value of the Dirac phase, oscillation experiments are going to perform simultaneous fits to neutrino and antineutrino data. In experiments employing the calorimetric method of energy reconstruction, neutrons---known to carry out larger fraction of the probe's energy in antineutrino interactions (see figure 3 in \cite{Ankowski:2015jya})---are an additional source of systematic uncertainty, as they typically deposit only part of their energy in the detector. Traveling some distance from the primary interaction point before scattering, neutrons are also currently problematic to associate with the neutrino event that knocked them out of the nucleus. In near detectors, much smaller than the far ones, they are likely to escape detection altogether. However, ongoing experimental efforts~\cite{Berns:2013usa,Liu:2015fiy} are expected to bring important progress in the understanding of the detector response to neutrons and reduce the related systematic uncertainties.

To emphasize the importance of reduction of systematic uncertainties in appearance measurements, let us make a simple estimate of those needed for an experiment to probe CP violation at 3$\sigma$ confidence level for 75\% of the possible $\delta_{CP}$ values, the goal set by the P5 advisory panel~\cite{P5}. Let us consider the CP asymmetry defined as~\cite{Benhar:2015wva,Huber:2014nga}
\begin{equation}\label{eq:asymmetry}
\mathcal{A}=\frac{\langle P(\nu_\mu\rightarrow\nu_e)\rangle-\langle P(\bar\nu_\mu\rightarrow\bar\nu_e)\rangle}{\langle P(\nu_\mu\rightarrow\nu_e)\rangle+\langle P(\bar\nu_\mu\rightarrow\bar\nu_e)\rangle},
\label{eq:cpa}
\end{equation}
with the energy average $\langle\dots\rangle$ taken over the full width at half the first oscillation maximum for a uniform flux. Note that in vacuum $\mathcal{A}$ is proportional to $\sin\delta_{CP}$.

\begin{figure}
\centering
\includegraphics[width=0.8\textwidth]{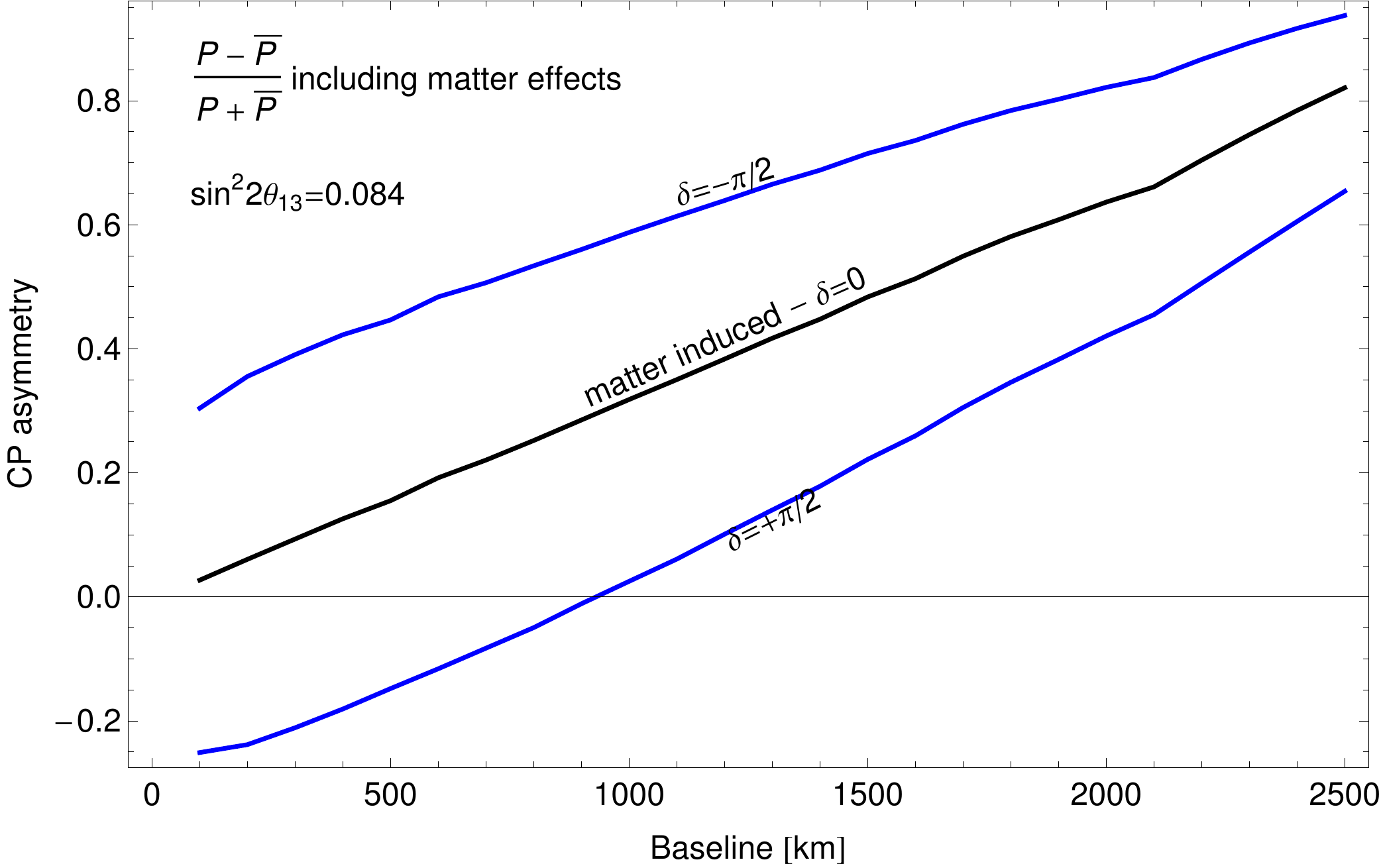}
\caption{\label{fig:cpa} Baseline dependence of the CP asymmetry~\eref{eq:asymmetry} for different $\delta_{CP}$ values calculated assuming the normal mass hierarchy~\cite{Benhar:2015wva,Huber:2014nga}.}
\end{figure}

\Fref{fig:cpa} shows the CP asymmetry $\mathcal{A}$ as a function of baseline length for the $\delta_{CP}$ values of $-\pi/2$, 0, and $+\pi/2$ and the remaining oscillation parameters from a recent global fit~\cite{GonzalezGarcia:2012sz}, assuming the normal mass hierarchy. For a 1500-km baseline, matter effects introduce the asymmetry of $\sim$47\% and $\delta_{CP}$ can modify it by $\pm25\%$.
For 75\% of the $\delta_{CP}$ values, the CP asymmetry can become as small as 5\%, which at 3$\sigma$ confidence level for CP violation discovery translates into $\sim$1.5\% accuracy of CP asymmetry measurement. For equal contributions of statistical and systematic uncertainties, the latter are required not to exceed $\sim$1\%. Fortunately, if the maximal CP violation is realized in nature, as suggested by the combination of the recent results of T2K~\cite{Abe:2015ibe} and NOvA~\cite{Adamson:2016xxw} with the reactor measurements~\cite{An:2015rpe,RENO:2015ksa,Abe:2014bwa}, less stringent conditions are going to be sufficient for the discovery of CP violation.

\section{Oscillation analysis}\label{sec:oscAnalysis}

Discussing the role of nuclear and detector effects in the oscillation analysis, in sections \ref{sec:nuclearEffects} and \ref{sec:detectorEffects} we present the results obtained using the software package \globes{}~\cite{Huber:2004ka,Huber:2007ji,Coloma:2012ji}. Simulating data according to~\eref{eq:rate}, we keep all the oscillation parameters fixed, setting their values to
\begin{equation}\label{eq:oscparams}
\eqalign{\theta_{13} =  9.0^\circ,\qquad \theta_{12} =  33.2^\circ,\qquad \theta_{23} =  45.0^\circ,\\
\Delta m^2_{21}  = 7.64\times 10^{-5} \, \textrm{eV}^2, \qquad \Delta m^2_{31}  = 2.45\times 10^{-3} \, \textrm{eV}^2.}
\end{equation}

Unless specified otherwise, we fix the Dirac phase to $\delta_{CP}=0$ and perform the $\nu_\mu\rightarrow\nu_\mu$ disappearance analysis for a setup similar---but due to simplifications not identical---to that of the T2K experiment~\cite{Abe:2011ks} with a narrow-band off-axis beam peaked at $\sim$600 MeV~\cite{Huber:2009cw}. In most of the cases, we consider carbon as the target material in the near and far detectors, to be able to assess uncertainties of the cross sections employed in the analysis by comparisons with available experimental data. Note that for oxygen and argon this would not be currently possible. The rationale for considering the T2K-like kinematics is its similarity to both Hyper-Kamiokande~\cite{Abe:2015zbg} and the region of the second oscillation maximum in DUNE~\cite{Acciarri:2015uup}, much more sensitive to $\delta_{CP}$ than the first one.

The main motivation for selecting the $\nu_\mu \rightarrow \nu_\mu$ channel is that it can effectively be treated as two-flavor oscillations with the probability~\cite{Nunokawa:2005nx}
\[
P_{\mu\mu} \simeq 1 - \sin^2\theta_{\mu\mu}\sin^2\left(\frac{ \Delta m^2_{\mu\mu}L}{4E_\nu}\right),
\]
where
\[
\eqalign{
\sin^2\theta_{\mu\mu} = 4\cos^2\theta_{13}\sin^2\theta_{23}(1 - \cos^2\theta_{13}\sin^2\theta_{23}),\\
\Delta m^2_{\mu\mu}=\Delta m^2_{31} + \Or(\Delta m^2_{21}),}
\]
which allows for a clear interpretation of the obtained results.

\begin{table}
\caption{\label{tab:setup}Experimental setup assumed for the disappearance analysis. The employed T2K-like neutrino flux~\cite{Huber:2009cw} is peaked at energy $\sim$600 MeV and has width of $\sim$200 MeV.}
\begin{indented}
\item[]\begin{tabular}{@{}lrr}
\br
 		& Baseline  	& Fiducial mass 	 \\
\mr
Far detector    	&  295.0~km    	& 22.5~kt \\
Near detector      	&  1.0~km     	& 1.0~kt  \\
\br
\end{tabular}
\end{indented}
\end{table}

As detailed in \tref{tab:setup}, the experimental setup consists of the near and far detectors simulated following~\cite{Huber:2009cw}. We do not take into account any differences in the beam acceptance or particle efficiencies of the detectors, which may be relevant in a real experiment. The exposure is assumed to be at least 5 years at the beam intensity 750 kW to ensure that the statistics of unoscillated CC events in the far detector exceeds 1000, sufficient to constrain the analyzed systematic uncertainty.

Our treatment of systematic uncertainties, the $\chi^2$ implementation and the method of determination of the confidence regions are the same as in \cite{Coloma:2012ji,Coloma:2013rqa,Coloma:2013tba}. The final $\chi^2$ is obtained after the minimization over the nuisance parameters $\xi$, as
\[
\chi^2=\min_\xi\left\{\sum_{D,\,i}\chi^2_{D,\,i}(\Delta m,\theta,\xi)+\sum_i\left(\frac{\xi_{\phi,\,i}}{\sigma_\phi}\right)^2+\left(\frac{\xi_N}{\sigma_N}\right)^2\right\},
\]
where the last two contributions are pull terms associated with the shape and the overall normalization, respectively. The shape parameters $\xi_{\phi,\,i}$ are bin-to-bin uncorrelated, and the normalization parameter $\xi_N$ is fully correlated between bins. The prior uncertainties $\sigma_\phi$ and $\sigma_N$ are set to 20\%, consistently with the spread between the available experimental data~\cite{Lyubushkin:2008pe,AguilarArevalo:2010zc,Fiorentini:2013ezn}. Note, however, that due to correlations between the near and far detectors in our analysis, our results turn out not to be very sensitive to this choice. For each energy bin $i$ and detector $D$, the contribution to $\chi^2$ is
\[
\chi^2_{D,\,i}(\Delta m,\theta,\xi)=2\left(F_{D,\,i}-O_{D,\,i}+O_{D,\,i}\ln\frac{O_{D,\,i}}{F_{D,\,i}}\right)
\]
where $O_{D,\,i}$ and $F_{D,\,i}$ denote the observed and fitted event rates in an energy bin $i$ and a detector $D$. While $O_{D,\,i}$ depends only on the assumed true values of the oscillation parameters, $F_{D,\,i}$ is a function of the fitted oscillation parameters and the nuisance parameters.


As neutral-current background is expected not to play an important role in the disappearance analysis, we neglect it---unless specified otherwise---and consider the following mechanisms of charged-current (CC) neutrino interaction: QE scattering involving a single nucleon or more nucleons (\ph{}) in the final state, resonance excitation and deep-inelastic scattering. In all cases, the events for resonance excitation and deep-inelastic scattering are obtained using the Monte Carlo generator \genie{} 2.8.0~\cite{Andreopoulos:2009rq}. On the other hand, QE events are simulated using either the relativistic Fermi gas (RFG) model from \genie{} or the realistic spectral function (SF) approach~\cite{Benhar:1989aw,Benhar:1994hw} implemented in the $\nu T$ package of additional modules~\cite{Jen:2014aja}. Note that the implementation of the SF approach in the $\nu T$ package~\cite{Jen:2014aja} differs from the one in \genie{} 2.8.0 by ensuring conservation of energy and momentum, and has been validated through comparisons to electron-scattering data for targets such as carbon, oxygen, argon and calcium.

For the \ph{} contribution, we employ the empirical procedure developed by Dytman~\cite{Katori:2013eoa} or---when specified---the effective approach of increasing the value of the axial mass $M_A$ in the QE cross sections. While these phenomenological methods are able to describe some aspects of experimental cross sections~\cite{Katori:2013eoa,Ankowski:2012ei}, it is clear that they cannot reliably predict neither composition not detailed kinematic distributions of the nucleonic final states. However, before the efforts toward developing an \emph{ab initio} estimate of \ph{} processes for neutrinos are successfully completed~\cite{Benhar:2015ula,Rocco:2015cil}, the two phenomenological approaches applied here seem to be sufficient to demonstrate how the existing spread between the available experimental data translates into an effect on the oscillation analysis relying on the lepton kinematics to reconstruct the neutrino energy.

Performing the kinematic energy reconstruction~\cite{Ahn:2002up}, we use the well-known formula
\begin{equation}\label{eq:kinERec}
E_\nu^\mathrm{kin}=\frac{2(M-\epsilon) E_\ell+W^2-(M-\epsilon)^2-m_\ell^2}{2(M-\epsilon-E_\ell+\n{k_\ell}\cos\theta)},
\end{equation}
setting the invariant hadronic mass $W$ to the nucleon mass $M$ for mesonless events and fixing the separation energy $\epsilon$ to 34 MeV. The same value of $\epsilon$ is added for every detected nucleon in the calorimetric method,
\begin{equation}\label{eq:calEnergy}
E^\mathrm{cal}_\nu=E_{\ell}+\sum_{i}(T^N_i+\epsilon)+\sum_{j}E_j,
\end{equation}
in which the neutrino energy is reconstructed summing the charged-lepton's energy $E_{\ell}=\sqrt{m_\ell^2 + {\bf k}_\ell^2}$, the kinetic energies of
the knocked-out nucleons $T^N_i$ and the total energy of any other particle produced $E_{j}$.

In an idealized scenario in which nuclear and detector effects would not play a significant role, the energy reconstruction would be perfect. In a realistic case, however, there is a non-negligible probability of energy misreconstruction \cite{Benhar:2009wi,Leitner:2010kp,Martini:2012fa,Nieves:2012yz,Meloni:2012fq,Lalakulich:2012hs,Martini:2012uc,Coloma:2013rqa,
Coloma:2013tba,Mosel:2013fxa,Jen:2014aja,Ankowski:2014yfa,Ankowski:2015jya,Ankowski:2015kya,Ericson:2016yjn,Ankowski:2016bji}.
In our calculations, both nuclear and detector effects are encoded in the migration matrices, $\mathcal{M}^X_{ij}$, the elements of which describe the probability that an event of type $X$ with a true energy in the bin $j$ is reconstructed with an energy in the bin $i$. The simulated (true and fitted) event distributions are then obtained as
\[
N^\mathrm{tot}_{i} = \sum_{X} \sum_{j} \mathcal{M}^{X}_{ij} N^{X}_j,
\]
where $X$ runs over the four types of interactions considered, $i$ and $j$ label the energy bins and $N^X_j$ stands for the true number of $X$ events in the bin $j$. The migration matrices used to obtain the results discussed in this review are provided as plots or in tabulated form in \cite{Coloma:2013tba,Jen:2014aja,Ankowski:2015jya,Ankowski:2016bji}.

\section{Nuclear effects}\label{sec:nuclearEffects}

As the oscillation parameters are extracted from the energy distribution of collected events, a precise reconstruction of neutrino energy is a prerequisite for a precise oscillation analysis. It is important to note that the kinematic method of energy re{\-}con{\-}struc{\-}tion~\eref{eq:kinERec}---valid for particular final states---relies on a selection of event subsample and requires an accurate simulation of both the signal events---QE ones in the past and ongoing experiments---and the background contribution~\cite{Coloma:2013rqa}. Analyzing various aspects of QE interactions with one or more nucleons in the final state~\cite{Coloma:2013tba,Jen:2014aja,Ankowski:2016bji}, in this section we discuss how uncertainties in the description of nuclear effects may affect the outcome of $\nu_\mu$ disappearance experiments employing the kinematic method of energy reconstruction. Although as a case study we consider an experimental setup similar to that of T2K, one needs to keep in mind that it is highly idealized and that our analysis method is simplified with respect to that employed by the T2K Collaboration. Here we do not take into account any detector effects.

While modern neutrino-scattering measurements have been predominantly per{\-}formed for the carbon target, the future oscillation experiments are going to require detailed knowledge of the cross sections for oxygen and argon. Until they are avail{\-}able, the oscillation studies are going to rely on nuclear models to extrapolate existing experimental results---predominantly for carbon---to the nuclei and kinematics of interest, as it is currently done e.g. in T2K~\cite{Abe:2016aoo}.
To demonstrate the subtleties of extrapolating the results between nuclei, we are going to discuss the importance of accounting for differences between the migration matrices for carbon ($A=12$) and oxygen ($A=16$) in the $\nu_\mu$ disappearance analysis.

Following Coloma \etal\cite{Coloma:2013tba}, let us consider a T2K-like experimental setup with wa{\-}ter-Cherenkov near and far detectors, and model nuclear effects relying exclusively on \genie{}. In particular, in this analysis QE scattering is described within the RFG model of Bodek and Ritchie~\cite{Bodek:1980ar}, with a high-momentum tail---inspired by the effect of nucleon-nucleon correlations---added to the nucleon momentum distribution given by the step function. The \globes{} analysis for 5 years of collecting data with the 750 kW beam~\cite{Huber:2009cw} yields $\sim$880 CC QE events, $\sim$480 pionless events resulting from other mechanisms of interaction and 275 neutral-current background events in the far detector.

\begin{figure}
\centering
    \includegraphics[width=0.39\textwidth]{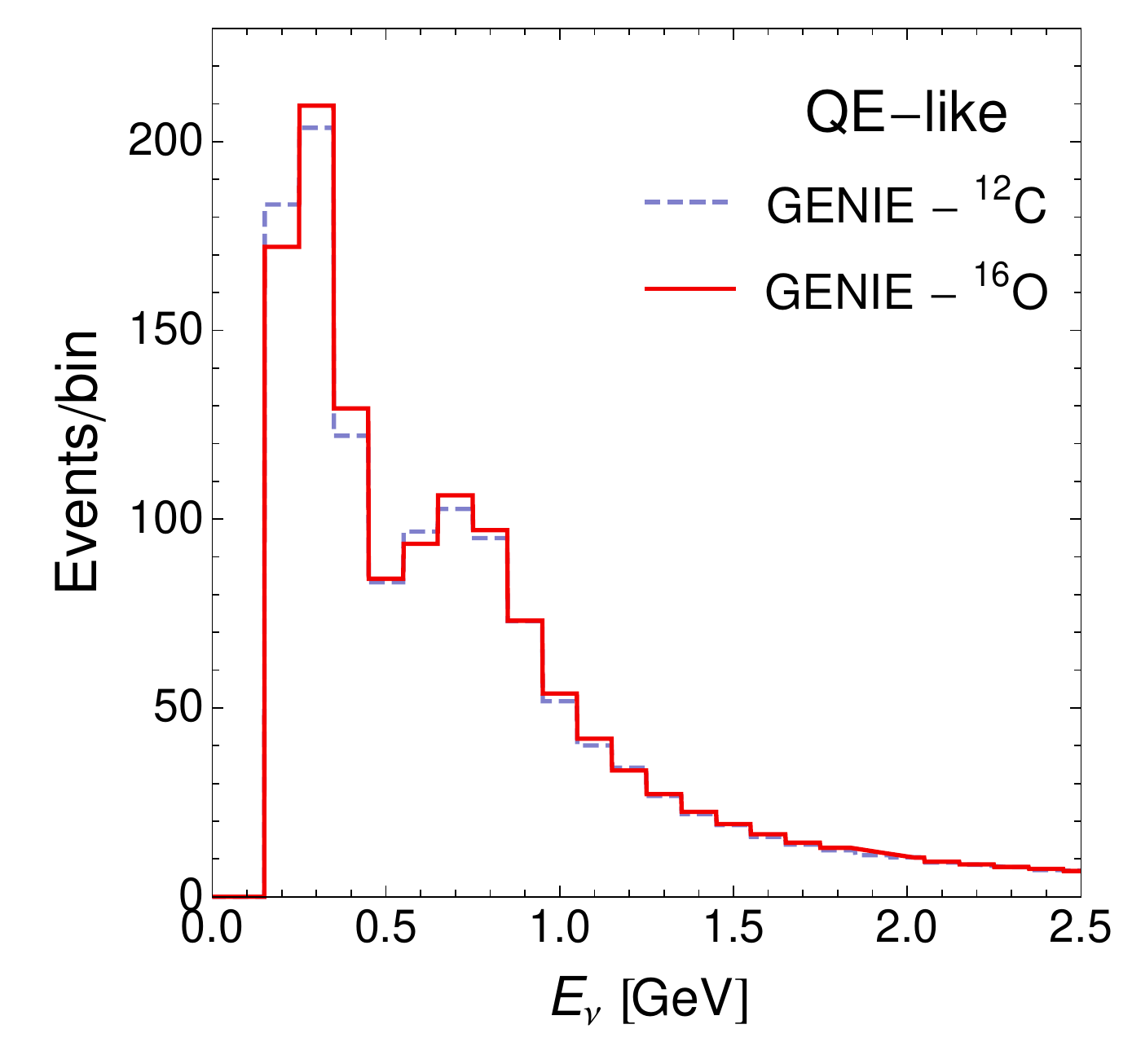}
    \hspace{0.8cm}
    \includegraphics[width=0.39\textwidth]{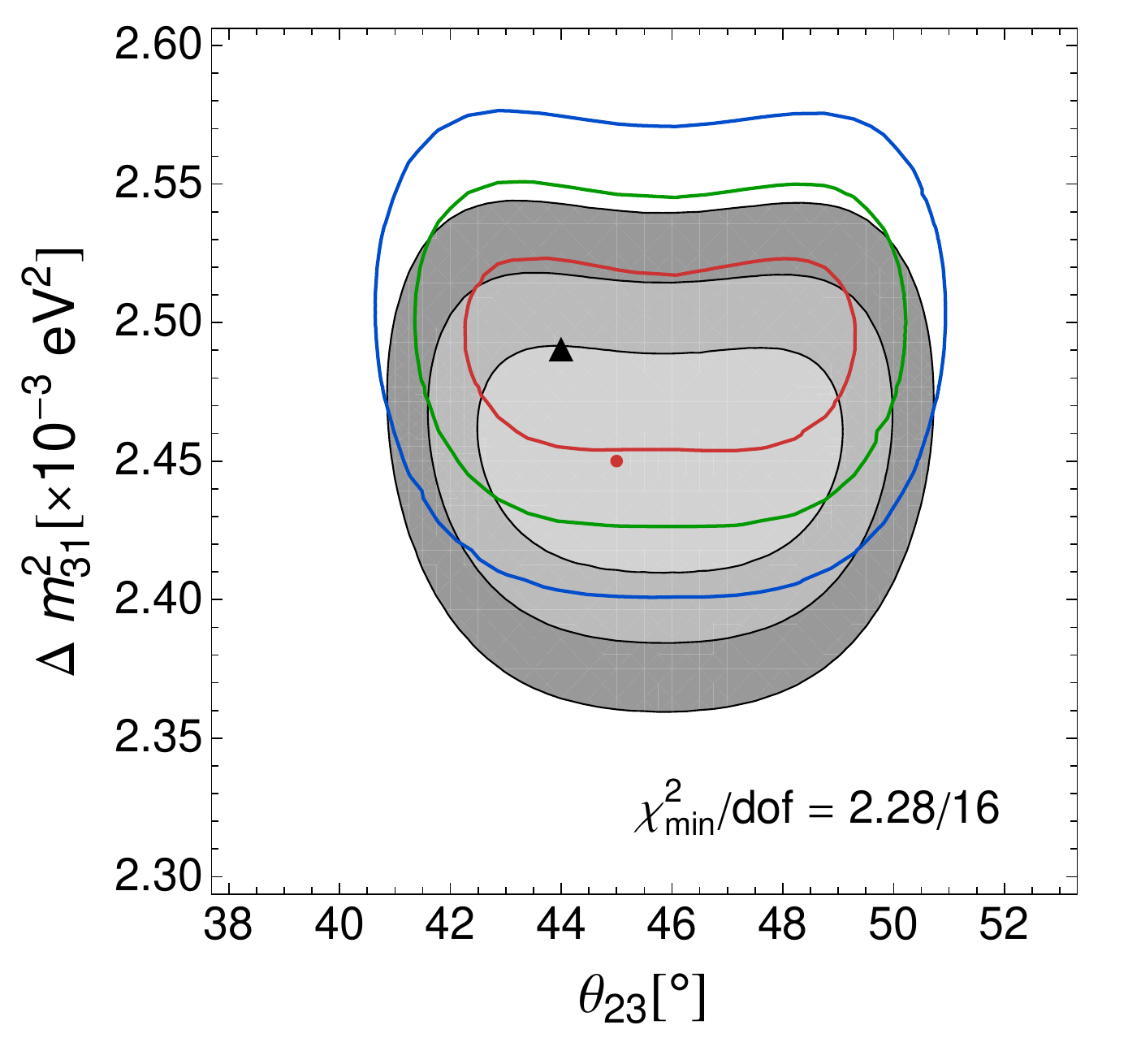}
\caption{\label{fig:CvsO}Impact of the nuclear target dependence of the cross sections in \genie{} on the oscillation analysis.
Left: Event distributions for oxygen (solid line) and carbon (dashed line) expected for the flux~\cite{Huber:2009cw}. Right: 1, 2 and 3$\sigma$ confidence regions in the $(\theta_{23},\,\Delta m_{31}^2)$ plane obtained when the simulated data for oxygen are fitted using the migration matrices for carbon (solid lines) and oxygen (shaded areas). The triangle and circle show the best fit point and true values of the oscillation parameters, respectively. Reprinted figure with permission from~\cite{Coloma:2013tba}, copyright 2014 by the American Physical Society.}
\end{figure}

The $E_\nu^\mathrm{kin}$ distributions of events obtained using the oxygen and carbon migration matrices are shown in the left panel of \fref{fig:CvsO}. The carbon result turns out to be somewhat shifted toward lower energies with respect to the oxygen one, predominantly due to differences in the CC QE cross sections, which translate into a noticeable effect for the oscillation analysis.

The right panel of \fref{fig:CvsO} shows the 1, 2 and 3$\sigma$ confidence regions in the $(\theta_{23}, \Delta m^2_{31})$ plane for the true event rates simulated using the oxygen migration matrices. The shaded areas represent the analysis in which nuclear effects are accurately taken into account by using the oxygen migration matrices to calculate the fitted rates. On the other hand, the solid lines correspond to the fitted rates for the carbon migration matrices and show the effect of an inaccurate description of nuclear effects. The extracted and true values of the oscillation parameters turn out to differ at a $\sim$1$\sigma$ confidence level.

We observe a non-negligible difference between the energy reconstruction for the carbon and oxygen targets described by means of the RFG model in \genie{}. Should the differences in their shell structures and density distributions be accounted for within a more sophisticated model, the effect can be expected to be even larger. Note also that carbon and oxygen are considered rather similar nuclei compared to argon.

Obviously, in a real experiment employing different targets in the near and far detectors, the difference between nuclear effects would be corrected for in the oscillation analysis using a nuclear model. As this procedure would be a source of uncertainties, the result of \fref{fig:CvsO} shows the importance of developing an accurate theoretical description of nuclear effects for the targets used in oscillation experiments and of their extensive testing against scattering data. It also suggests that employing the same target in the near and far detectors can be an effective way to reduce systematic uncertainties of an oscillation experiment.

It is important to note that when the interaction process involves only one nucleon---with the remaining $(A-1)$ ones acting as spectators---neutrino and electron scattering are subject to the same nuclear effects. The nuclear cross section in this regime is a convolution of the elementary cross section (different for electrons and neutrinos) with the hole and particle spectral functions (common for electrons and neutrinos), describing the ground-state properties of the target and  propagation of the struck nucleon, respectively~\cite{Benhar:2006wy}. As a consequence, models of \emph{nuclear effects} in neutrino interactions can be validated by systematic comparisons to electron-scattering data at the kinematics of interest. We want to emphasize that it is highly improbable that a theoretical approach unable to reproduce electron-scattering data would be able to describe nuclear effects in neutrino interactions. Moreover, electron-scattering cross sections---double differential and for monoenergetic beams---allow for much easier understanding of discrepancies between theoretical results and experimental data than neutrino ones, in which different probe's energies and reaction mechanisms are intertwined.

\begin{figure}
\centering
\includegraphics[width=0.46\columnwidth]{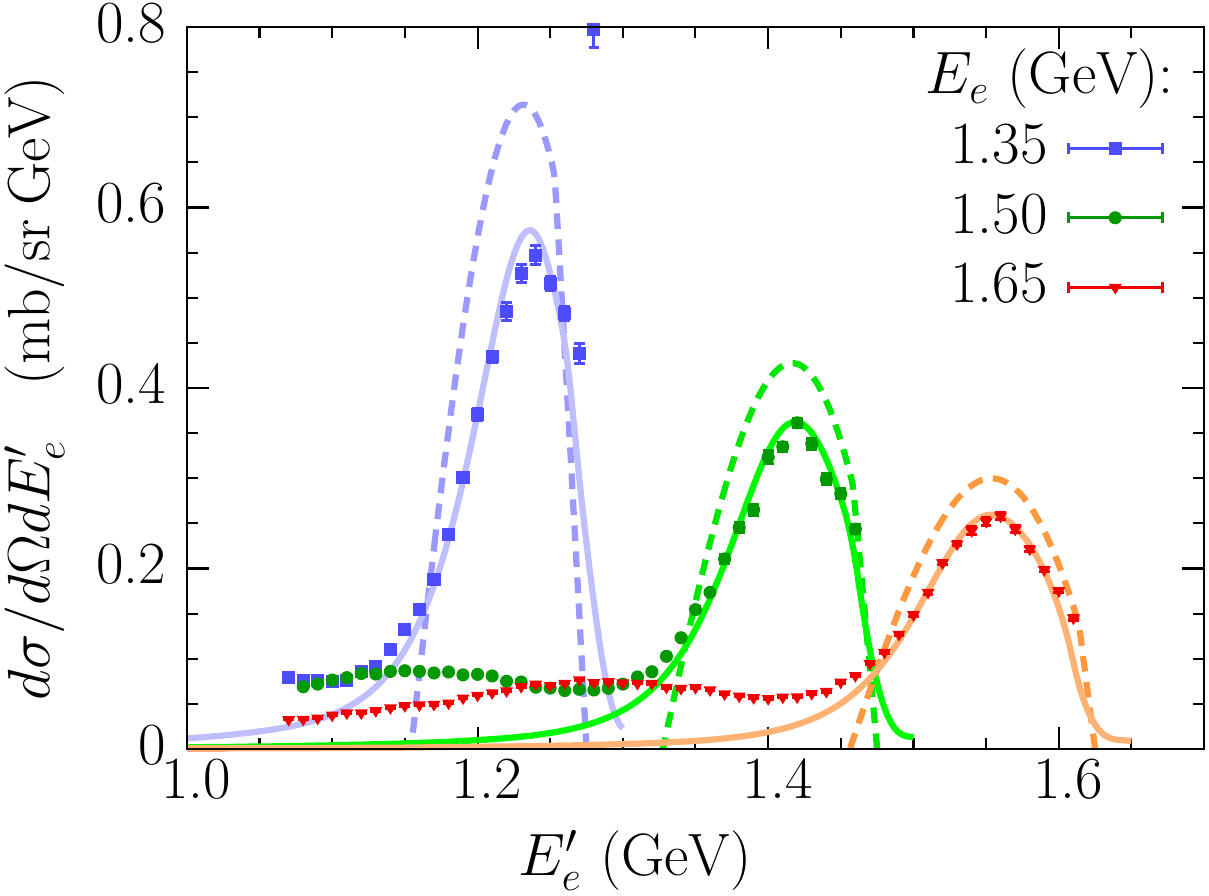}
\hspace{0.8cm}
\includegraphics[width=0.39\columnwidth]{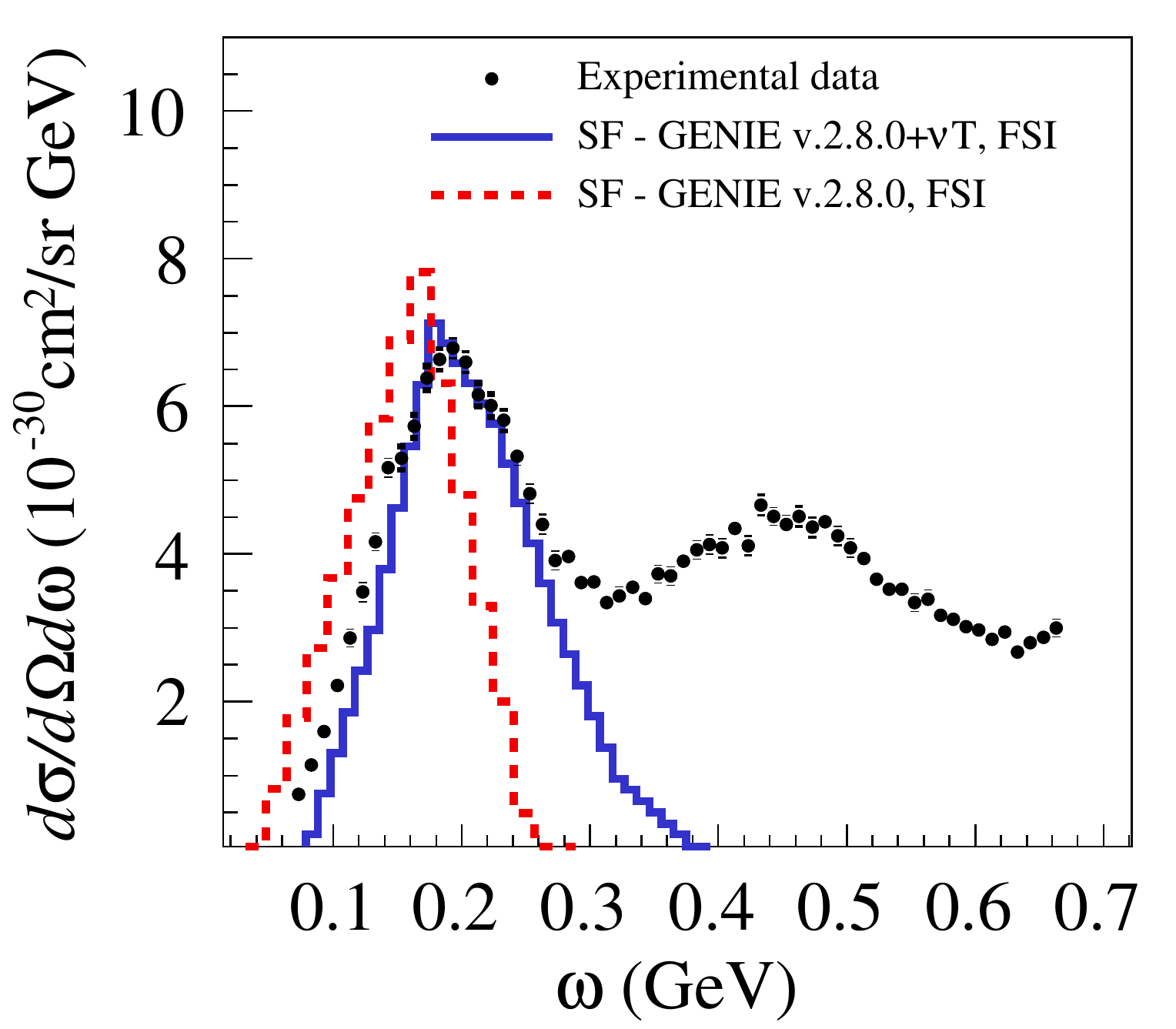}
\caption{\label{fig:electrons}Importance of an accurate description of nuclear effects on the example of the double differential C$(e,e')$ cross sections. Left: Comparison of the RFG (dashed lines) and  SF~\cite{Ankowski:2014yfa} (solid lines) calculations with the data for scattering angle $13.5^\circ$ and beam energies 1.35, 1.50 and 1.65 GeV~\cite{Baran:1988tw}. Right: Comparison of the SF results obtained using \genie{}~\cite{Andreopoulos:2009rq} and \vt{}~\cite{Jen:2014aja} with the experimental data for scattering angle $37.5^\circ$ and beam energy 961 MeV \cite{Sealock:1989nx}.}
\end{figure}

As an illustrative example, in the left panel of \fref{fig:electrons} we compare the QE predictions of the SF calculations~\cite{Ankowski:2014yfa} and the RFG model of Smith and Moniz~\cite{Smith:1972xh} with the data~\cite{Baran:1988tw}. Whereas the RFG results cannot reproduce heights, shapes and positions of the QE peaks, the SF calculations---although they do not involve any adjustable parameters---turn out to be in very good agreement with the experimental points.

Using the realistic ground-state description~\cite{Benhar:1989aw,Benhar:1994hw}, the SF calculations take into account both the shell structure and correlations between nucleons. The latter cause partial depletion of the shells---lowering the QE peaks---and give rise to high-momentum nucleons deeply bound in quasideuteron pairs, producing the tails at low values of the final electron energy $E_e'$ and rendering the QE peaks asymmetric. In addition, the SF calculations~\cite{Ankowski:2014yfa} include the effect of final-state interactions, essential to reproduce the positions of the QE peaks accurately. All these effects are neglected in the RFG model, which treat the nucleus as a fragment of noninteracting nuclear matter of uniform density
in a constant potential.

The SF approach is available in \genie{} 2.8.0, however, comparisons with electron-scattering data have revealed some issues with the energy and momentum conservations in its implementation that affect both the position and width of the QE peak, as shown in the right panel of \fref{fig:electrons} for the experimental cross sections~\cite{Sealock:1989nx}. On the other hand, the SF implementation in the $\nu T$ package~\cite{Jen:2014aja} of additional modules to \genie{} is in good agreement with electron-scattering data and, therefore, from now on we are going to rely on it to generate QE events whenever we discuss the SF approach.

\begin{figure}
\centering
\includegraphics[trim=0 -14 0 0bp,width=0.53\columnwidth]{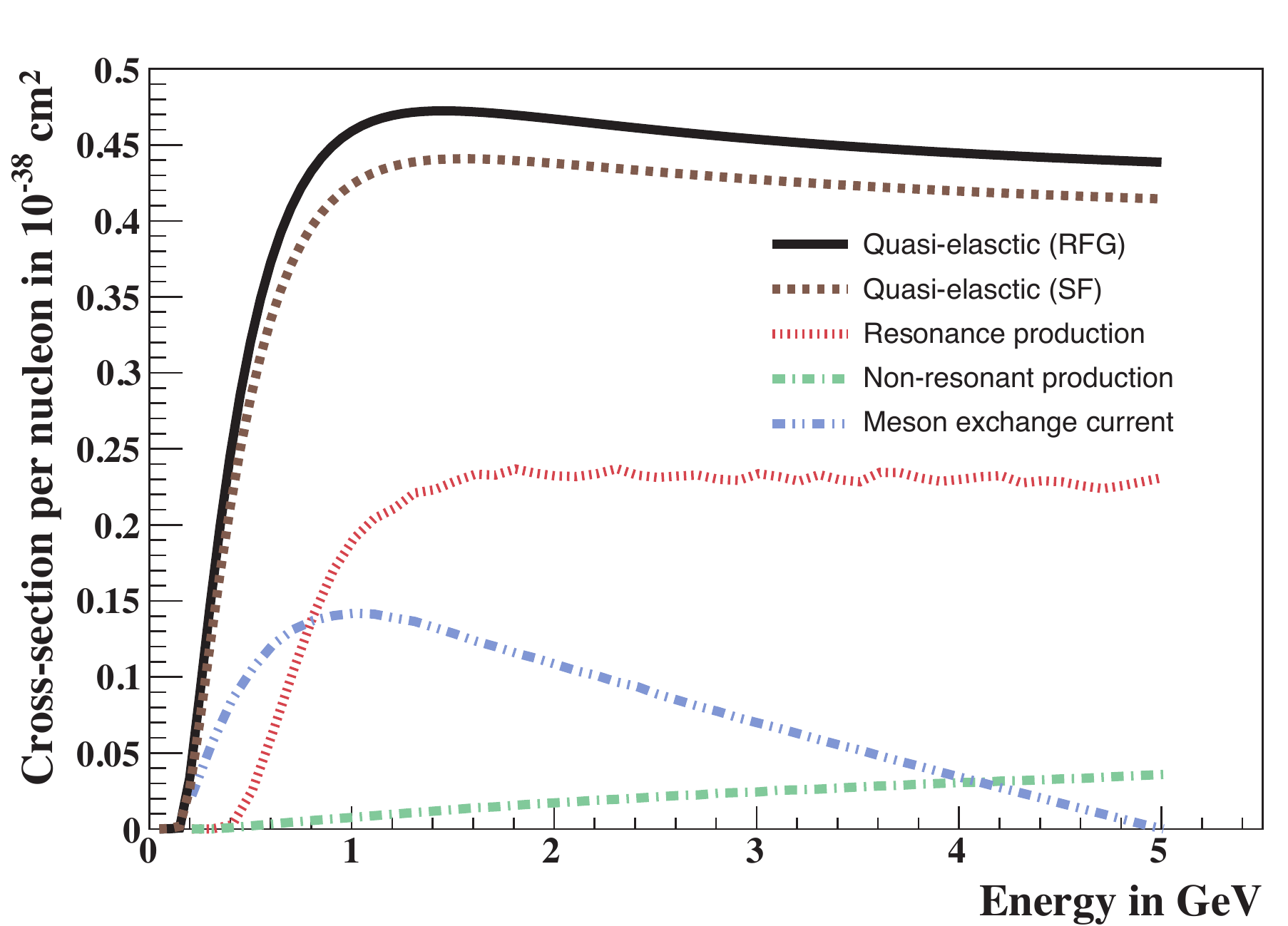}
\includegraphics[width=0.39\columnwidth]{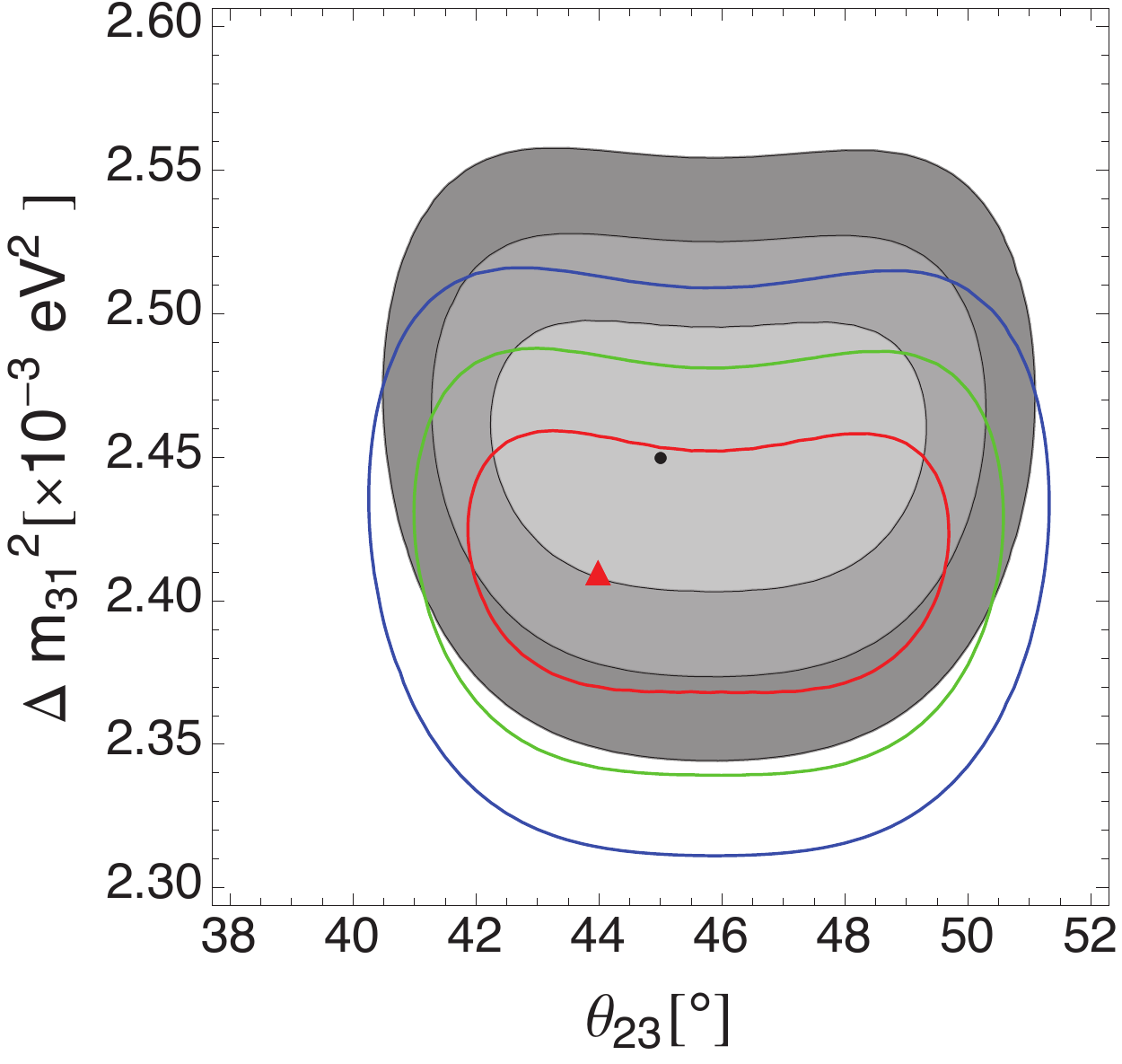}
\caption{\label{fig:SFvsRFG}Impact of differences between the quasielastic \nuC{} cross sections in the SF approach and the RFG model on the oscillation analysis. Left: Mesonless cross sections employed in the analysis. Right: Confidence regions in the $(\theta_{23},\,\Delta m_{31}^2)$ plane obtained when the data simulated within the SF approach are fitted using the migration matrices from the RFG (solid lines) and SF (shaded areas) calculations.
Reprinted figure with permission from~\cite{Jen:2014aja}, copyright 2014 by the American Physical Society.
}
\end{figure}

To explore how the difference between the RFG model and the SF approach in the QE interaction channel influences the oscillation analysis~\cite{Jen:2014aja}, let us consider a $\nu_\mu$ disappearance experiment similar to T2K, employing carbon as the target material in the near and far detectors. Assuming 5 years of data taking at the beam power of 750 kW in the \globes{} analysis leads to $\sim$650 ($\sim$730) CC QE events in the SF approach (RFG model), together with $\sim$410 CC events of QE topology coming from other interaction mechanisms and $\sim$250 neutral-current background events. The difference between the CC QE event numbers in the two considered approaches, resulting from the difference between the corresponding cross sections presented in the left panel of \fref{fig:SFvsRFG}, contributes to a non-negligible effect for the oscillation analysis.

The right panel of \fref{fig:SFvsRFG} shows the 1, 2 and 3$\sigma$ confidence regions in the $(\theta_{23}, \Delta m^2_{31})$ plane for the true event rates simulated using the SF migration matrix for CC QE events. The shaded areas represent the fitted rates obtained using the same matrix, while the solid lines correspond to the RFG migration matrix. We observe a shift between the true values of the oscillation parameters---represented by the dot---and the extracted best-fit point---marked by the triangle---at a $\sim$1$\sigma$ confidence level, both in the mixing angle and the squared mass difference.


As we analyze the difference between the RFG model and the SF approach only for the CC QE predictions, the presented analysis should be extended to consistently take into account multinucleon processes~\cite{Benhar:2015ula,Rocco:2015cil} and final-state interactions~\cite{Ankowski:2014yfa} in order to provide a complete quantitative estimate of the influence of nuclear-model uncertainties. Moreover, for resonance excitation, similar differences between the RFG model and the SF approach can be expected based on the available theoretical results~\cite{Benhar:2006nr} as long as the same form factors enter the calculations. Note also that in the deep-inelastic region, the current modeling of nuclear effects has been shown inaccurate by the recent results from the MINERvA experiment~\cite{Tice:2014pgu}.

From now on, we rely on the SF approach to simulate CC QE events and employ a generalization of the standard formula~\eref{eq:kinERec} for the kinematic energy reconstruction~\cite{Ankowski:2015jya}: in the case of mesonless events, we employ the invariant hadronic mass $W=M$, regardless of the number of knocked-out nucleons, but when at least one meson is detected in the final state, we set $W$ to the $\Delta$ resonance mass $M_\Delta=1.232$ GeV.

To discuss the effect of uncertainties of the \ph{} cross sections on the oscillation analysis, we consider a T2K-like disappearance experiment running in the neutrino and antineutrino mode with the same flux~\cite{Huber:2009cw} and employing the carbon target in the near and far detectors. We compare the results obtained using two data-driven phenomenological methods to account for \ph{} processes~\cite{Ankowski:2016bji}. In the \vt{} approach,
the \ph{} estimate from \genie{} 2.8.0~\cite{Katori:2013eoa} is added to the QE contributions obtained using the $\nu T$ package~\cite{Jen:2014aja}. In the effective approach, the modifications of the QE cross sections due to \ph{} mechanisms are accounted for by applying in the SF calculations an effective value of the axial mass $M_A = 1.2$ GeV, as suggested by
a number of experimental results for nuclear targets ranging from carbon to iron~\cite{Gran:2006jn,AguilarArevalo:2007ab,Mariani:2008zz,AguilarArevalo:2010zc,Abe:2014iza,Adamson:2014pgc}. The rationale for considering these two methods is that the spread of the cross sections they predict is a good representation of the spread between the available experimental data. Note, however, that as these methods are purely phenomenological, they are unlikely to accurately predict the number of knocked-out nucleons or details of their kinematics.

\begin{figure*}
\centering
{\includegraphics[width=0.47\textwidth]{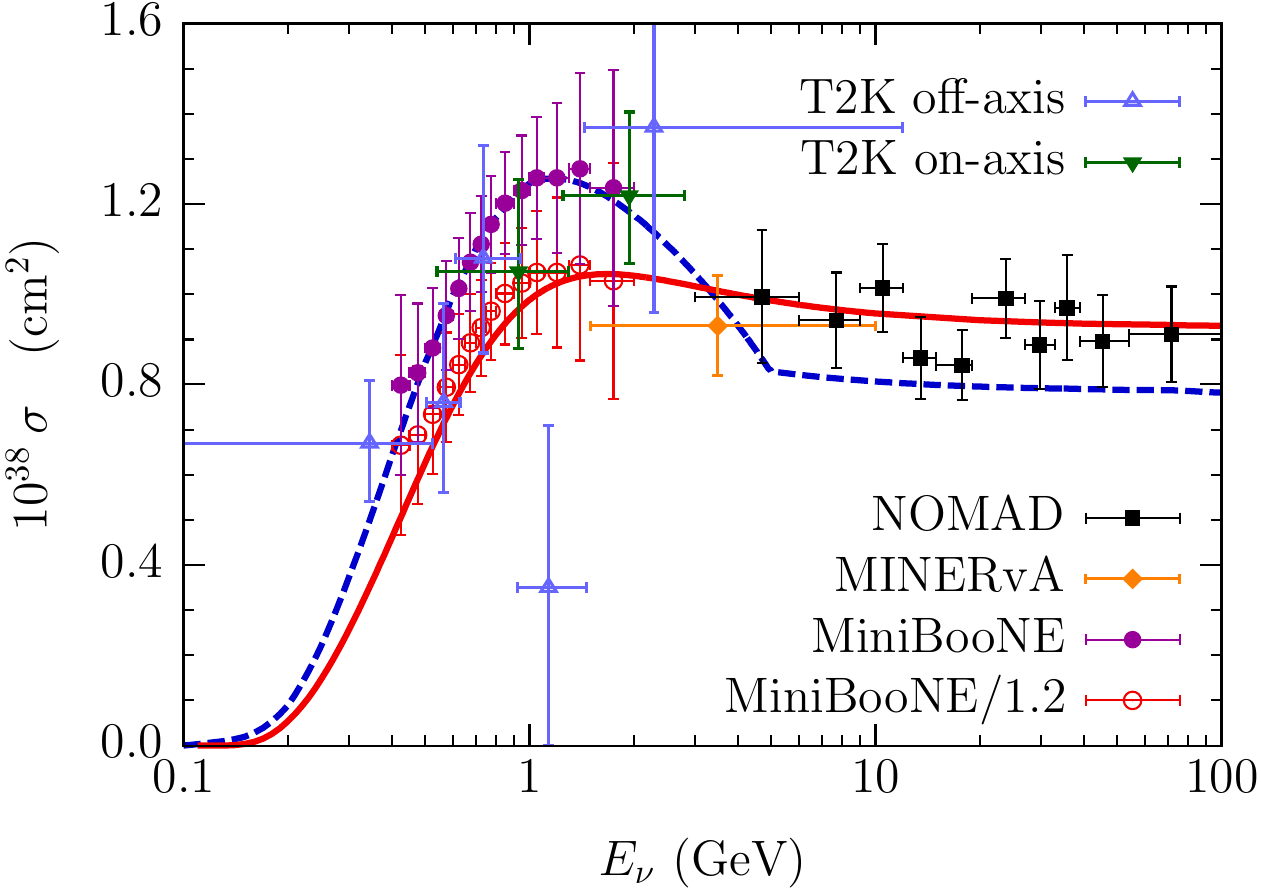}}
    \hspace{0.4cm}
{\includegraphics[width=0.47\textwidth]{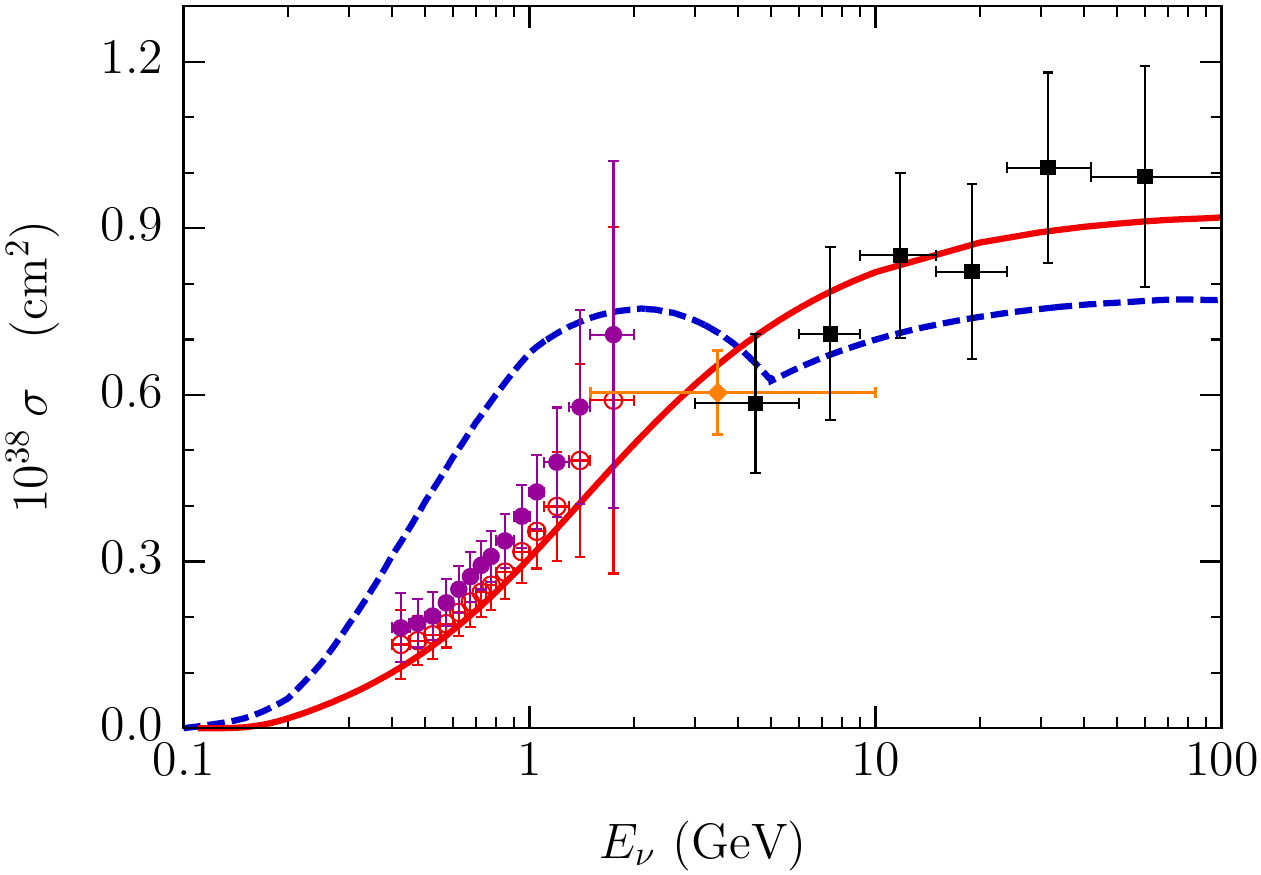}}
\caption{\label{fig:xsec} Charged-current quasielastic $\nu_\mu$ (left) and $\bar\nu_\mu$ (right) cross sections. The \vt{} (dashed lines) and effective (solid line) calculations for carbon are compared with the carbon data from NOMAD~\cite{Lyubushkin:2008pe} and MiniBooNE~\cite{AguilarArevalo:2010zc,Aguilar-Arevalo:2013dva} and the hydrocarbon measurements from MINERvA~\cite{Fiorentini:2013ezn,Fields:2013zhk} and T2K~\cite{Abe:2015oar,Abe:2014iza}. For comparison, we also show the MiniBooNE data divided by 1.2. Reprinted figure with permission from~\cite{Ankowski:2016bji}, copyright 2016 by the American Physical Society.}
\end{figure*}

\Fref{fig:xsec} shows that the effective calculations are in good agreement with the CC QE results from NOMAD~\cite{Lyubushkin:2008pe} and MINERvA~\cite{Fields:2013zhk,Fiorentini:2013ezn}, for both neutrinos and antineutrinos. They also reproduce the energy dependence of the neutrino and antineutrino data from MiniBooNE~\cite{AguilarArevalo:2010zc,Aguilar-Arevalo:2013dva}, but not their absolute normalization. To better illustrate this feature, we also present the MiniBooNE cross sections divided by a factor of 1.2, consistent with the ratio of the detected to predicted events $1.21\pm0.24$ in the first MiniBooNE analysis~\cite{AguilarArevalo:2007ab}.
The \vt{} approach is in very good agreement with the MiniBooNE data for neutrinos, which were used to determine the strength of the \ph{} contribution in \genie{}~\cite{Katori:2013eoa}, but this is not the case for antineutrinos. Owing to their large uncertainties, the T2K data~\cite{Abe:2015oar,Abe:2014iza} cannot discriminate between the two calculations.

As a side remark, we note that the understanding of CC interactions without pions in the final state has undergone an important improvement in the past few years and a non-negligible role of reaction mechanisms involving more than one nucleon is now generally acknowledged. Therefore, comparisons with older measurements should be taken with a pinch of salt, having also in mind that some of them involved model dependent correction for the phase space not covered by the detectors. In ongoing experiments great care is taken to report results with minimal model dependence, as a function of muon kinematics.

\begin{figure}
\centering
    {\includegraphics[width=0.53\textwidth]{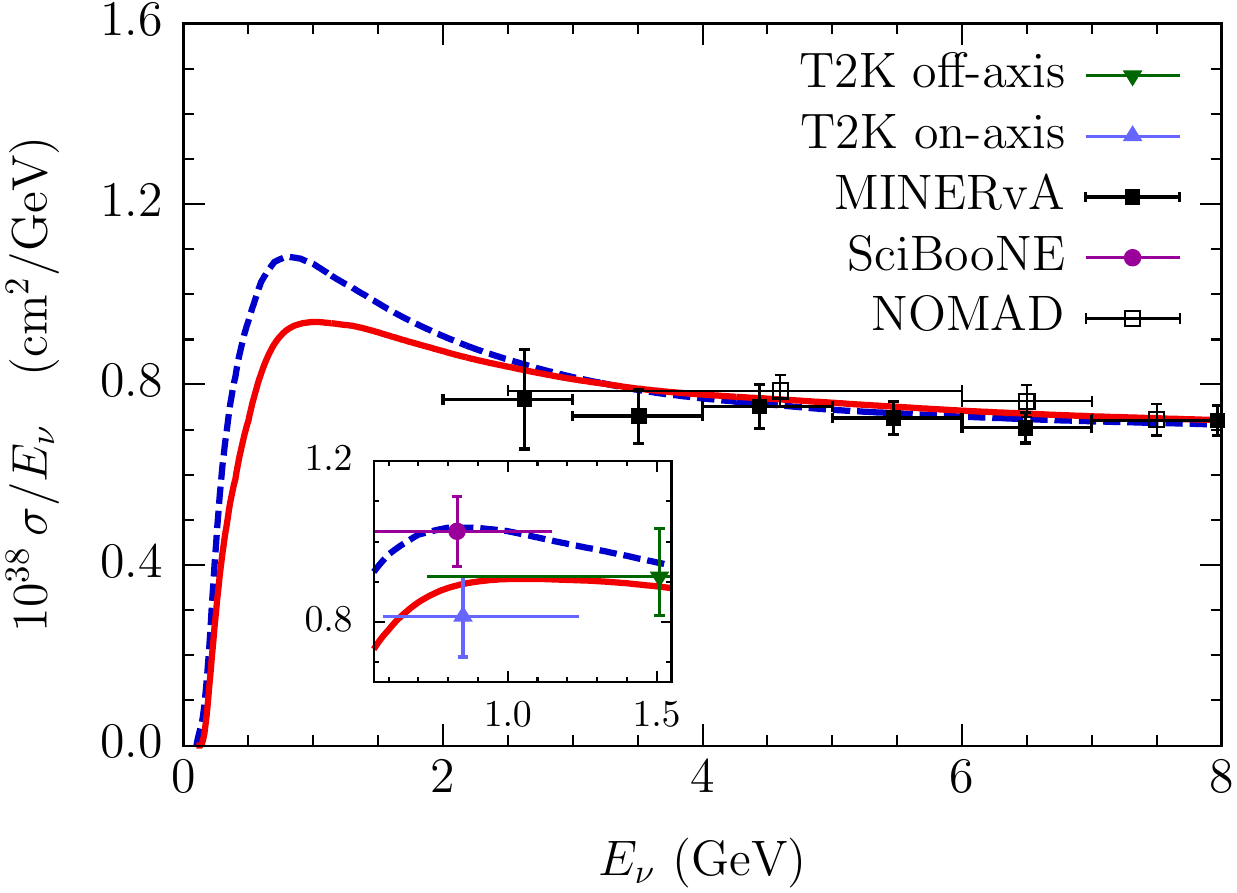}}
    \hspace{0.8cm}
    {\includegraphics[width=0.39\textwidth]{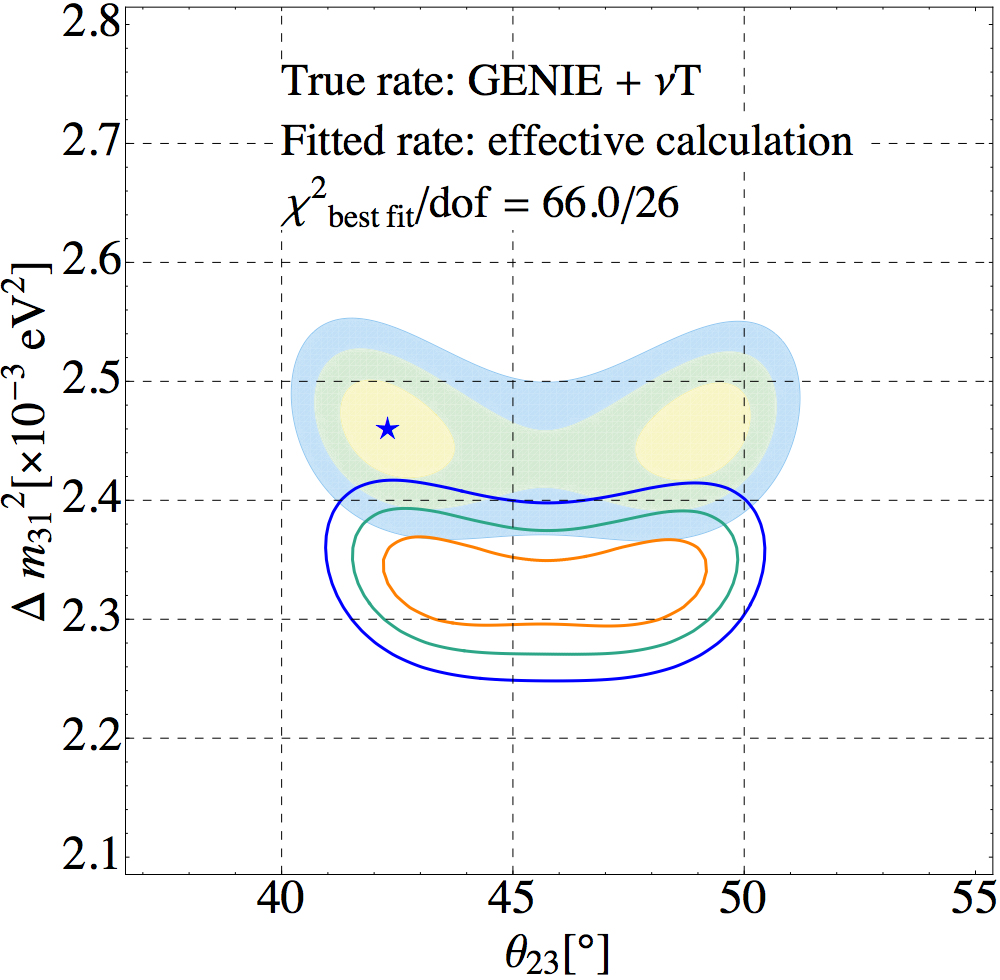}}
\caption{\label{fig:2p2h_nu}Impact of uncertainties of the $2p2h$ cross section for muon neutrinos on the oscillation analysis. Left: Inclusive \nuCX{} cross sections obtained using the effective (solid line) and \vt{} (dashed line) calculations are compared with the NOMAD~\cite{Wu:2007ab} and MINERvA~\cite{DeVan:2016rkm} data. The inset presents the hydrocarbon results and flux-averaged measurements reported by the SciBooNE~\cite{Nakajima:2010fp} and T2K~\cite{Abe:2013jth,Abe:2014nox} Collaborations. Right: 1, 2 and 3$\sigma$ confidence regions in the $(\theta_{23}, \Delta m^2_{31})$ plane obtained when the data simulated within the \vt{} approach are fitted using the migration matrices from the effective (solid lines) and \vt{} (shaded areas) calculations. The star shows the true values of the oscillation parameters. Reprinted figure with permission from~\cite{Ankowski:2016bji}, copyright 2016 by the American Physical Society.
}
\end{figure}

Comparison of the total CC inclusive $\nu_\mu$ cross sections in the left panel of \fref{fig:2p2h_nu} shows that both the \vt{} approach and the effective calculations yield the results in good agreement with the NOMAD~\cite{Wu:2007ab} and MINERvA~\cite{DeVan:2016rkm} data, collected in the region dominated by deep-inelastic scattering.
While the flux-averaged on-axis measurement from T2K~\cite{Abe:2014nox} does not distinguish the two approaches, the off-axis one~\cite{Abe:2013jth} shows a distinct preference for the effective calculations. On the other hand, the SciBooNE point~\cite{Nakajima:2010fp} clearly favors the \vt{} calculations. Note that the T2K~\cite{Abe:2013jth,Abe:2014nox} and SciBooNE~\cite{Nakajima:2010fp} measurements do not involve energy unfolding or selection of particular event topologies and, therefore, they can be expected to involve the lowest uncertainties.

The puzzling difference between the T2K and SciBooNE data has important consequences for $\nu_\mu$ disappearance studies. To illustrate it, in the right panel of \fref{fig:2p2h_nu} we show the confidence regions for the true event rates obtained using the \vt{} migration matrices and the expected number of $\sim$6000 unoscillated events, which in the neutrino mode corresponds to $\sim$5 years of data collecting at the beam power 750 kW. The shaded areas and solid lines represent the results for the fitted rates calculated using the migration matrices from the \vt{} and effective calculations, respectively. The high value of $\chi^2$ per degree of freedom in the best-fit point, 66.0/26, clearly indicates that the differences between the two considered approaches are too large to be neglected in a precise oscillation analysis.

\begin{figure}
\begin{center}
\includegraphics[width=0.53\textwidth]{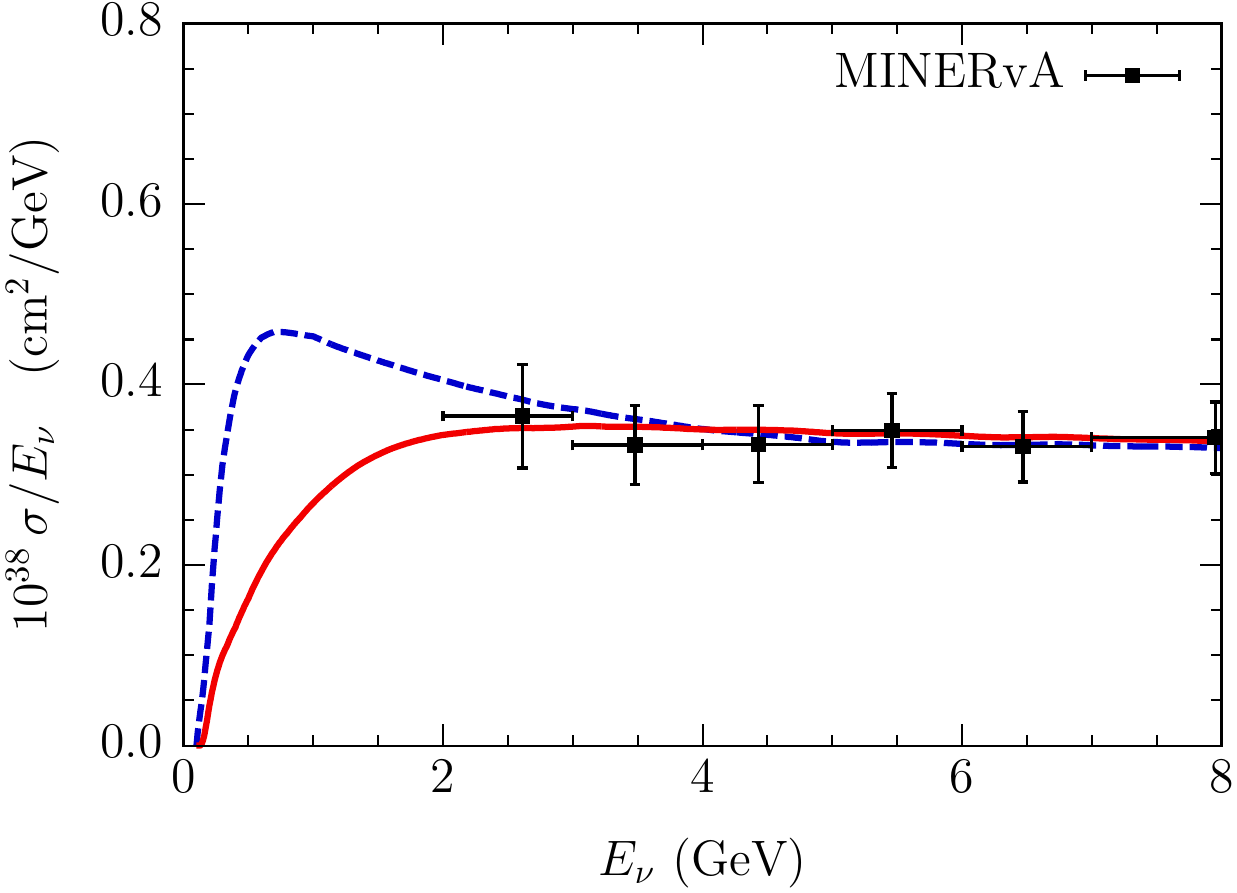}
\hspace{0.8cm}
\includegraphics[width=0.39\textwidth]{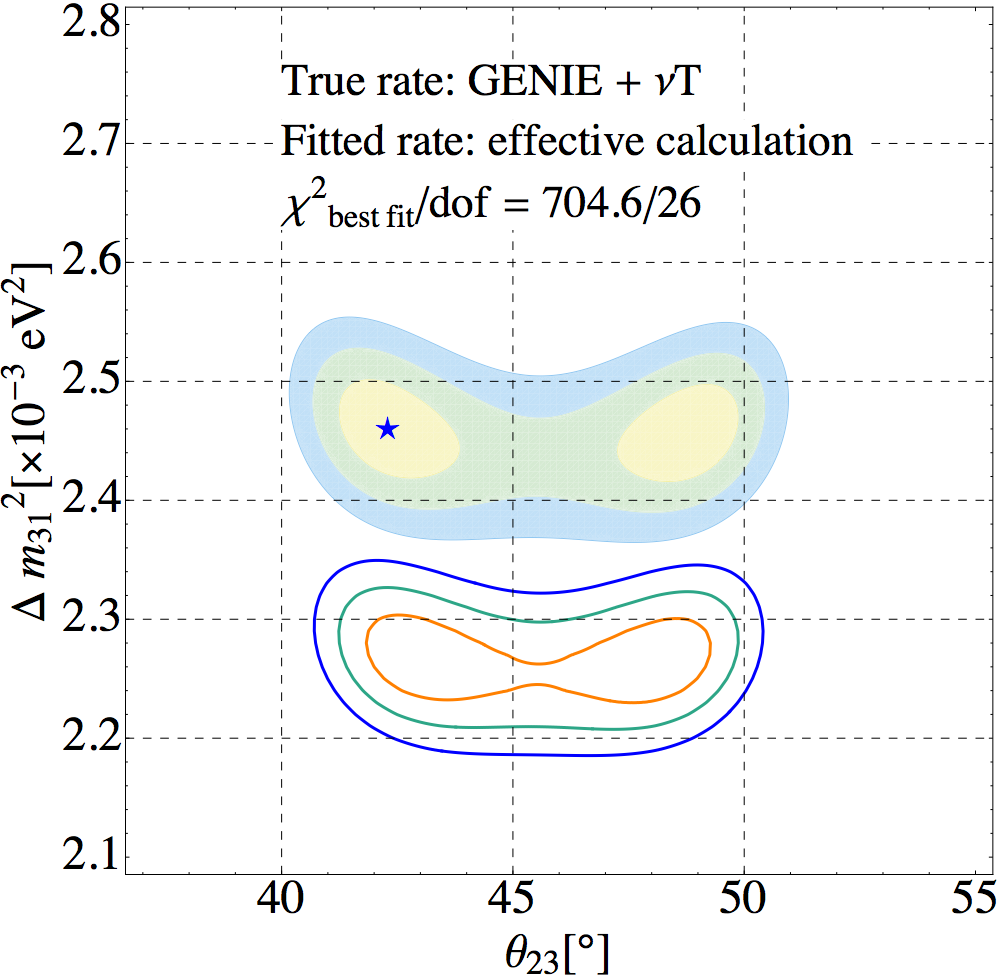}
\caption{\label{fig:2p2h_anu}Same as figure \ref{fig:2p2h_nu} but for muon antineutrinos. Reprinted figure with permission from~\cite{Ankowski:2016bji}, copyright 2016 by the American Physical Society.}
\end{center}
\end{figure}

For completeness, in \fref{fig:2p2h_anu} we present the results for muon antineutrinos. Owing to the large differences between the $\bar\nu_\mu$ CC QE cross sections in the low energy region, where experimental data are currently unavailable, the two considered approaches yield very different predictions for the inclusive CC $\bar\nu_\mu$ cross section, as shown in the left panel of \fref{fig:2p2h_anu}. On the other hand, at the kinematics probed by the MINERvA experiment~\cite{DeVan:2016rkm}, dominated by deep-inelastic scattering, they are consistent and in good agreement with the data.

The observed differences at low energies translate into a large effect for the oscillation analysis. Using the fitted rates from the effective calculations and the true event rates from \vt{} leads to a severe discrepancy between the extracted and true values of the oscillation parameters, see the right panel of \fref{fig:2p2h_anu}. Because treating the normalization of the QE event sample---with any number of nucleons---as arbitrary yields very similar results, the observed behavior can be traced back to the different shapes the two approaches predict  for the migration matrices and for the energy dependence of the CC QE cross section.

In final remarks of this section we note that the significance of nuclear effects, analyzed here for the carbon nucleus ($A=12$), can be expected to increase further for the argon target $(A=40)$. Although the considered experimental setup and our procedure of data analysis do not closely follow those of any existing experiment, the discussed results point toward the crucial role that an accurate description of nuclear effects plays in Monte Carlo simulations of high-statistics oscillation experiments. For the success of precise oscillation studies it is, therefore, of great importance to continue efforts aiming to provide new cross-section measurements, improve existing Monte Carlo generators, develop more accurate nuclear models and determine their uncertainties in comparisons with available data.

\section{Detector effects}\label{sec:detectorEffects}

In the oscillation analysis of long-baseline experiments, neutrino energy is inferred from the kinematics of particles produced in the interaction. As a consequence, the reconstructed energy may be affected by finite detection capabilities---energy resolutions, efficiencies and thresholds---and uncertainties they are known with.

We explore this issue within a realistic scenario, in which particles are detected according to their efficiencies and thresholds, and measured energies and angles are smeared by finite detector resolutions. To analyze how uncertainties of these detector effects may affect the oscillation analysis, we obtain simulated event distributions in the far detector within the realistic scenario and analyze them partly neglecting the detector effects implemented in the migration matrices.

Accounting for the effect of finite detector resolution, we smear observables according to the normal distributions centered at their true values. For muons, this procedure is applied both to the momentum and production angle, using the realistic parameters~\cite{Aliaga:2013uqz}
\begin{equation}\label{eq:muonResolutions}
\sigma(\n{k_\mu})=0.02\n{k_\mu}\quad\textrm{ and }\quad\sigma(\theta)=0.7^\circ.
\end{equation}
The energy resolutions of for $\pi^0$'s producing electromagnetic showers and other hadrons are set to
\begin{equation}\label{eq:hadResolutions}
\frac{\sigma(E_{\pi^0})}{E_{\pi^0}}=\max\left\{\frac{0.107}{\sqrt{E_{\pi^0}}},\, \frac{0.02}{E_{\pi^0}}\right\}\textrm{ and }
\frac{\sigma(E_h)}{E_h}=\max\left\{\frac{0.145}{\sqrt{E_h}},\,0.067\right\},
\end{equation}
respectively. Their values, as well as those of the energies appearing in the above equation, are expressed in units of GeV.

In our calculations, the detection thresholds correspond to the measured kinetic energy of 20 MeV for mesons and 40 MeV for protons. For the sake of simplicity, the efficiencies are treated as energy-independent and set to 60\% for $\pi^0$'s, 80\% for other mesons and 50\% for protons. Owing to difficulties of their reconstruction in neutrino events, we assume that neutrons always escape detection.

In the context of the $\nu_\mu$ disappearance analysis~\cite{Ankowski:2015jya}, we consider the T2K-like setup described in \sref{sec:oscAnalysis}, following the same treatment of systematic uncertainties and applying the same method of $\chi^2$ calculations. Our analysis is performed for the expected overall number of unoscillated events $\sim$4900, corresponding to $\sim$4 years of data collecting with the beam power 750 kW. We simulate the true event rates using the migration matrices calculated with detector effects.

Instead of varying individual parameters involved in the detector description, we vary the migration matrices making linear combinations of those calculated with and without detector effects. In this manner, we estimate the general sensitivity of the oscillation analysis to the detector performance, which can be expected to characterize a broad class of experiments rather than a particular (highly idealized) setup. These linear combinations of the migration matrices are then used to obtain the fitted event rates.

\begin{figure}
\begin{center}
\includegraphics[trim=0 -10 0 20bp,width=0.54\textwidth]{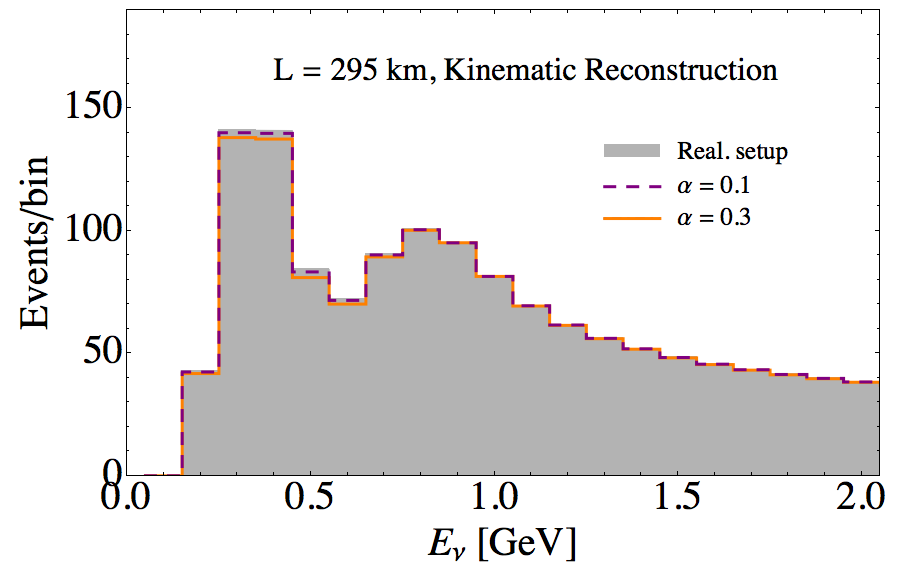}
\hspace{0.8cm}
\includegraphics[width=0.39\textwidth]{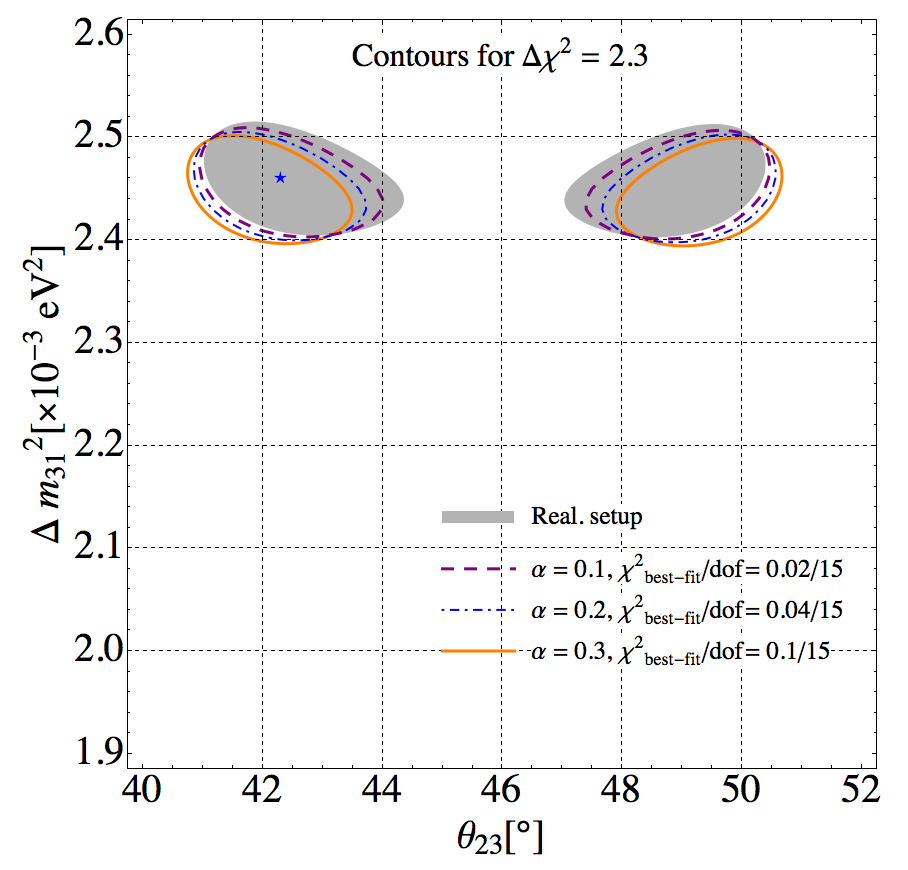}
\caption{\label{fig:detEff_kin}Impact of detector-related uncertainties on the oscillation analysis em{\-}ploy{\-}ing the kinematic energy reconstruction. Left: Comparison of the event distributions for the detector performance overestimated by 10\% (dashed line) and 30\% (solid line)  with an accurate estimate of detector effects (shaded histogram). Right: $1\sigma$ confidence regions in the $(\theta_{23},\,\Delta m_{31}^2)$ plane obtained when the data simulated with detector effects are fitted using the migration matrices neglecting them at the 10\%, 20\%, 30\% level (lines)---as shown by the value of $\alpha$---and accounting for them accurately (shaded area). Reprinted figure with permission from~\cite{Ankowski:2015jya}, copyright 2015 by the American Physical Society.}
\end{center}
\end{figure}

As shown in \fref{fig:detEff_kin}, the kinematic method of energy reconstruction---based only on the muon's energy and its production angle---turns out to be largely insensitive to detector effects and even the biases at the level of 30\% do not significantly affect the outcome of the oscillation analysis. It is a consequence of very precise reconstruction of muon kinematics achieved in modern experiments and assumed in this analysis, see~\eref{eq:muonResolutions}. Note, however, that our considerations do not take into account any issues related to the acceptance differences between the near end far detectors.

\begin{figure}
\begin{center}
\includegraphics[trim=0 -10 0 20bp,width=0.54\textwidth]{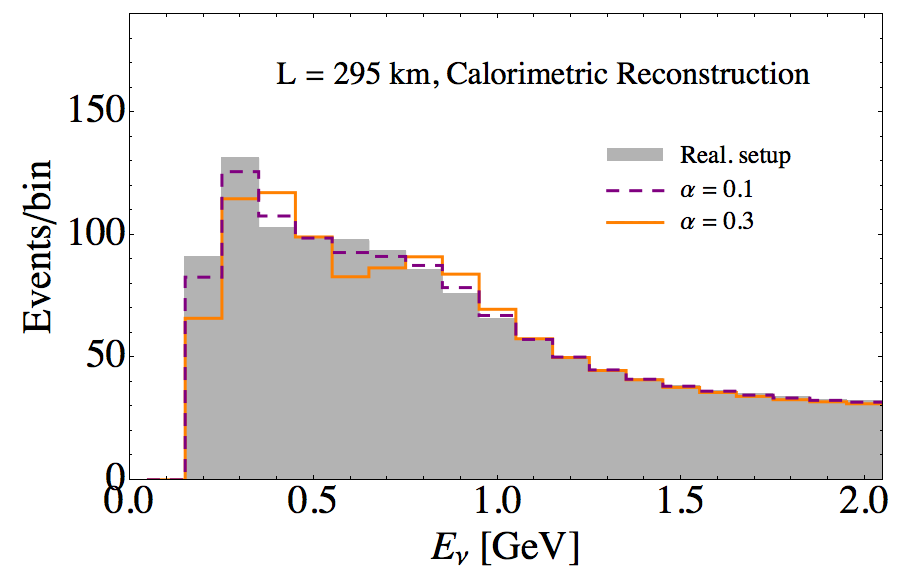}
\hspace{0.8cm}
\includegraphics[width=0.39\textwidth]{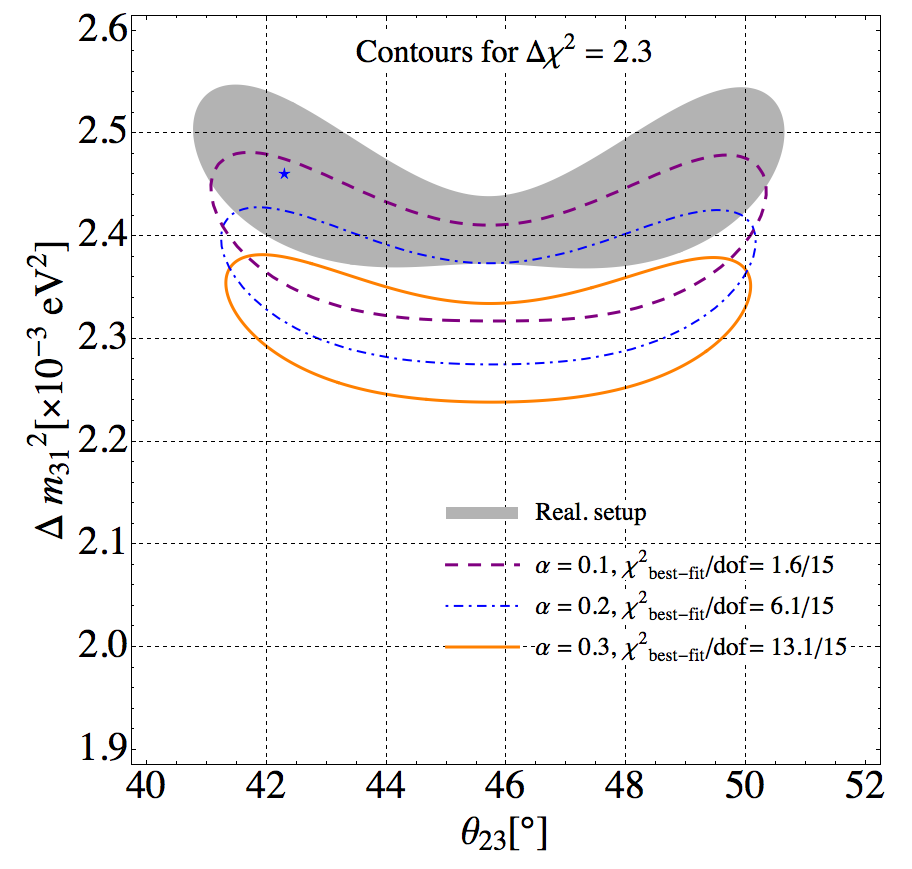}
\caption{\label{fig:detEff_cal}Same as figure \ref{fig:detEff_kin} but for the calorimetric energy reconstruction. Reprinted figure with permission from~\cite{Ankowski:2015jya}, copyright 2015 by the American Physical Society.}
\end{center}
\end{figure}

On the other hand, we observe that detector effects play an important role in the calorimetric reconstruction method, in which the neutrino energy is inferred from the energies deposited in detector by all interaction products. \Fref{fig:detEff_cal} shows that in order to avoid a significant bias in the extracted values of the oscillation parameters, the detector response has to be determined with an accuracy of at least 10\%. This behavior can be traced back to much larger uncertainties in the hadron-energy determination, see \eref{eq:hadResolutions}, entering the calorimetric method of neutrino energy reconstruction.

While the results of figures \ref{fig:detEff_kin} and \ref{fig:detEff_cal} are obtained for the narrow beam peaked at $\sim$0.6~GeV~\cite{Huber:2009cw}, the conclusions on influence of detector effects seem to be valid in more general case, as suggested by the findings for a~wide-band beam with the peak at $\sim$1.6 GeV and a non-negligible contribution of energies above 3 GeV~\cite{Ankowski:2015jya}.

Owing to the complicated oscillation probability and the necessity to disentangle parameter degeneracies~\cite{Nath:2015kjg}, an accurate neutrino-energy reconstruction is also im{\-}por{\-}tant for appearance experiments, observing electron (anti)neutrinos from oscil{\-}lations of muon (anti)neutrinos. In this context, we analyze the role of missing energy~\cite{Ankowski:2015kya}---carried away from event by undetected particles---on the $\delta_{CP}$ sensitivity for an experiment similar to DUNE~\cite{Acciarri:2016crz}.

In the considered setup, a wide-band neutrino beam is produced in interactions of the initial 1.08-MW proton beam with the target material and aimed at the far detector of fiducial mass 40 kton, located 1300 km from the source. We assume 6 years of collecting data, 3 in $\nu$ mode and 3 in $\bar\nu$ mode. For the signal, we consider 2\% uncertainties of normalization (bin-to-bin correlated) and shape (bin-to-bin uncorrelated). For the background, only a global normalization uncertainty of 5\% is taken into account~\cite{Ankowski:2015kya}.

The simulated event distributions are obtained accounting for all detector effects---resolutions, efficiencies and thresholds. On the other hand, extracting the oscillation parameters, we construct the migration matrices as linear combinations of those calculated with and without the shift resulting from the missing energy. In this way, for the purpose of this analysis, we single out the role of missing energy from other detector effects.

\begin{figure}
\begin{center}
\includegraphics[width=0.455\textwidth]{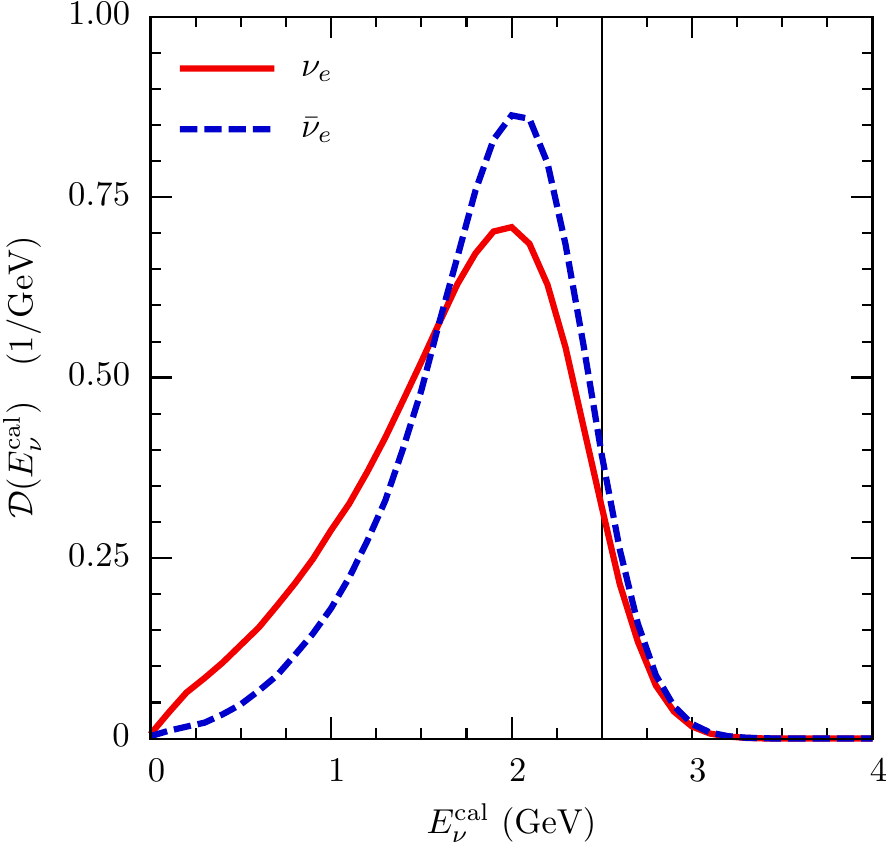}
\hspace{0.8cm}
\includegraphics[width=0.44\textwidth]{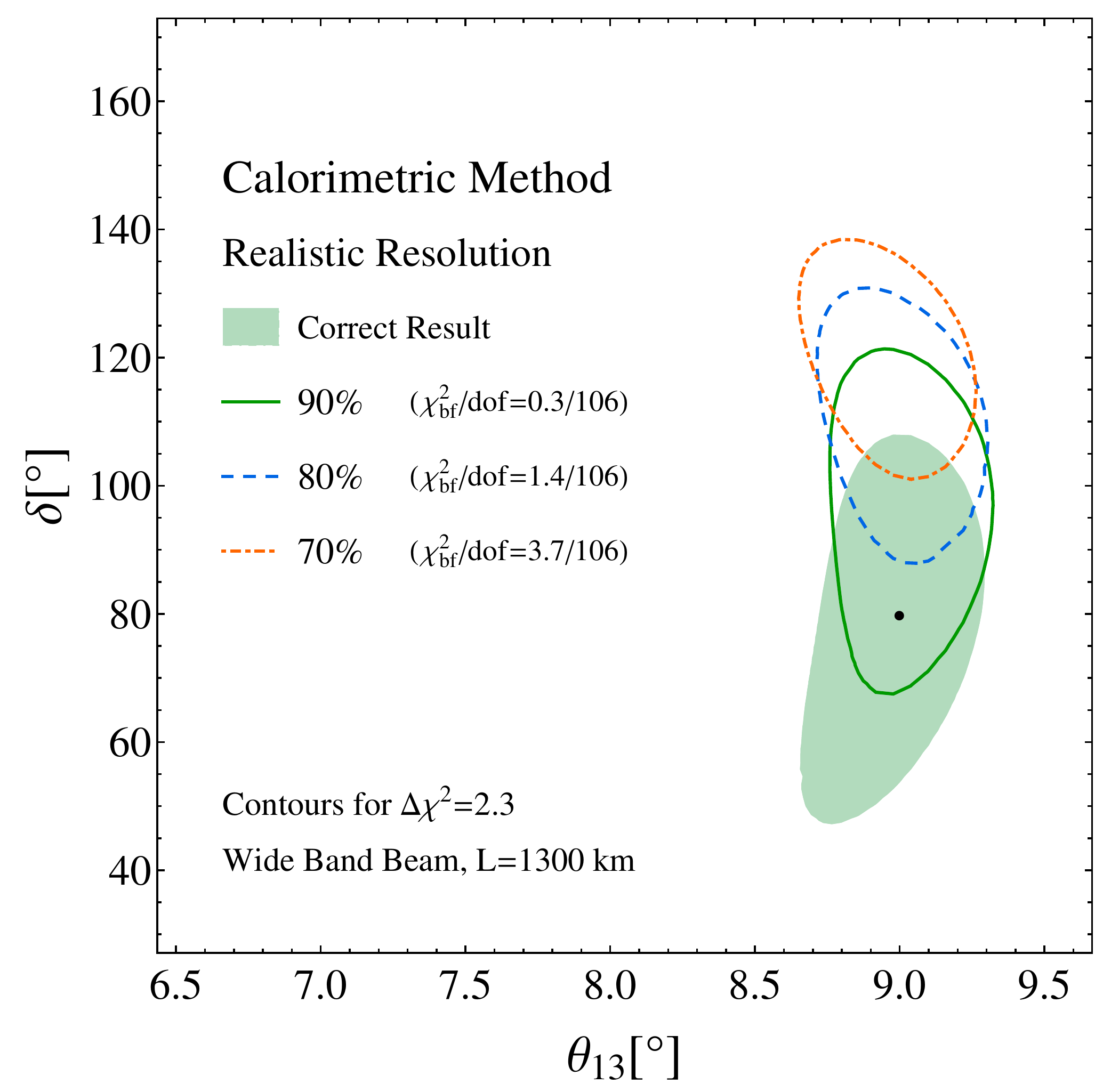}
\caption{\label{fig:missE}Impact of an underestimation of the missing energy on the oscil{\-}la{\-}tion analysis employing the calorimetric energy reconstruction. Left: Reconstructed energy distributions for deep-inelastic scattering of electron neutrinos and antineutrinos at true energy 2.5 GeV. Right: $1\sigma$ confidence regions in $(\theta_{13},\,\delta_{CP})$ plane obtained when the simulated data are fitted using the migration matrices accounting for 90\%, 80\% and 70\% (lines) of the missing energy and when all the missing energy is accounted for (shaded area). The dot shows the assumed true values of the oscillation parameters. Reprinted figure with permission from~\cite{Ankowski:2015kya}, copyright 2015 by the American Physical Society.}
\end{center}
\end{figure}

Its influence on the reconstructed-energy distributions is illustrated in the left panel of \fref{fig:missE} for the example of deep-inelastic scattering of electron neutrinos and antineutrinos at the true energy fixed to 2.5 GeV, at which this interaction channel contributes 37\% of the total inclusive cross section. Note that deep-inelastic scattering is the dominant mechanism of interaction at this kinematics---corresponding to the peak of the DUNE beam~\cite{Acciarri:2015uup}---and at higher energies. It clearly appears that owing to undetected particles in the final state, the maxima of the distributions are shifted to energies lower than the true energy, marked by the solid vertical line. As the composition of particles in the final state differs for neutrinos and antineutrinos, and depends on the interaction channel and the probe's energy, this is also the case for the energy carried away by undetected particles. In the deep-inelastic channel, we observe that the missing energy is on average $\sim$20\% lower for antineutrinos than for neutrinos. This feature largely stems from the destructive interference of the response functions in the antineutrino cross section at high energy transfers~\cite{Amaro:2004bs}, leading to the muon energy higher for antineutrinos than for neutrinos. Note, however, that for quasielastic scattering at $E_\nu=2.5$ GeV---contributing 25\% of the total inclusive cross section---the missing energy for antineutrinos is higher by $\sim$22\% than for neutrinos, due to the presence of undetected neutrons in the final state.

As presented in the right panel of \fref{fig:missE}, even a 20\% underestimation of the missing energy
introduces a sizable bias in the extracted $\delta_{CP}$ value. Should the missing energy be underestimated by 30\%, the analysis would exclude the true value of $\delta_{CP}$ at a confidence level between 2 and 3$\sigma$. This result illustrates the importance of an accurate determination of detector response in test-beam exposures and the relevance of a realistic simulation of nuclear effects in neutrino interactions, including intranuclear cascade. Because the differences between the missing energy for neutrinos and antineutrinos can be as large as $\sim$20\%, they can be expected to have important consequences for the oscillation analysis, and their neglecting could result in a $\sim$1$\sigma$ bias in the extracted $\delta_{CP}$ value.

We would like to emphasize that while our results are meant to point out the importance of various aspects of detector effects, much more detailed studies are necessary to draw truly quantitative conclusions for specific experiments.
In particular, our treatment of the missing-energy uncertainty---assumed to be equal for neutrinos and antineutrinos and independent of the energy and interaction channel---may be regarded as simplistic. However, as more realistic sensitivity estimates would require an accurate knowledge of the detector response and inclusion of nuclear-model uncertainties, out of necessity, we leave them for future investigations within experimental collaborations.

Finally, we note that while in our considerations detector effects have been analyzed separately from nuclear effects, in a real experiment they are intertwined and inaccuracies of their description cannot be  be properly disentangled and diagnosed.  The situation is even more involved when the near and far detectors qualitatively differ, because the detector effects in the far detector cannot then be constrained by the near-detector data.
 These problems increase the importance of relying on an accurate description of nuclear effects and determining the response of both the near and far detectors in extensive exposures to a variety of test beams.

\section{Summary}\label{sec:summary}
Thanks to tremendous progress in experimental neutrino physics over the last two decades, various systematic uncertainties in oscillation studies have been greatly reduced. As a consequence, however, those arising from description of nuclear effects and related to determination of detector response have become very important in current appearance measurements. This is going to be even more so in future experiments aiming to unambiguously discover the lepton-sector contribution to violation of charge-particle symmetry.

In this review, we have discussed a few examples illustrating the relevance of an accurate description of quasielastic interactions with any number of knocked-out nucleons and argued that for pion production a similar effect is expected. We pointed out that comparisons to electron-scattering data provide opportunity to test nuclear models and estimate their uncertainties; they also allow for much clearer interpretation of discrepancies than neutrino data. 
Discussing detector effects, we have drawn attention to their larger importance for experiments employing the calorimetric method of energy reconstruction and emphasized the key role of an accurate estimate of the missing energy for precise oscillation studies.

As a final remark we reiterate that future experiments are going to require co{\-}or{\-}di{\-}nat{\-}ed efforts toward improving theoretical descriptions of nuclear effects, implementing them in Monte Carlo generators and determining detector response in extensive test-beam exposures. The success of these efforts is a prerequisite for the discoveries of the next two decades, which however have the potential to overshadow those of the past two decades.

\section*{Acknowledgments}
We would like to express our gratitude to Omar Benhar, Pilar Coloma, Patrick Huber, Chun-Min Jen, Leonidas Kalousis, Davide Meloni and Erica Vagnoni for their contributions to the results discussed in this review.
This work is supported by the National Science Foundation under Grant PHY-1352106.

\end{document}